\documentclass[sigplan,nonacm]{acmart}

\usepackage{subcaption, algorithm, algorithmic, float, multirow, makecell, xspace, ragged2e}
\usepackage[normalem]{ulem}

\usepackage{adjustbox}
\usepackage{array}
\usepackage{tabularx}
\usepackage{booktabs}
\usepackage{tikz}

\newcommand{\yes}[0]{\tikz\draw[black,fill=black] (0,0) circle (.8ex);}
\newcommand{\sometimes}[0]{\tikz\draw[black,fill={gray!40}] (0,0) circle (.8ex);}
\newcommand{\no}[0]{\tikz\draw[black,fill=white] (0,0) circle (.8ex);}

\AtBeginDocument{
  \providecommand\BibTeX{{
    \normalfont B\kern-0.5em{\scshape i\kern-0.25em b}\kern-0.8em\TeX}}}

\begin{document}

\title{Slasher: Power Flexibility for Cloud Datacenters}

\author{\vspace{1pt}Liuzixuan Lin}
\affiliation{
  \institution{University of Chicago}
  \country{United States}
}

\author{\vspace{1pt}Fiodar Kazhamiaka}
\affiliation{
    \institution{Microsoft Azure}
    \country{United States}
}
\author{\vspace{1pt}Alok Gautam Khumbare}
\affiliation{
    \institution{Microsoft Azure}
    \country{United States}
}
\author{\vspace{1pt}Chaojie Zhang}
\affiliation{
    \institution{Microsoft Azure}
    \country{United States}
}
\author{\vspace{1pt}Jaylen Wang}
\affiliation{
    \institution{Carnegie Mellon University}
    \country{United States}
}
\author{\vspace{1pt}Hassan Khan}
\affiliation{
    \institution{Microsoft Azure}
    \country{United States}
}

\author{\vspace{1pt}Rodrigo L. Assis}
\affiliation{
    \institution{Microsoft Azure}
    \country{United States}
}

\author{\vspace{1pt}Mariana Rodrigues}
\affiliation{
    \institution{Microsoft Azure}
    \country{United States}
}

\author{\vspace{1pt}Kyle Woolcock}
\affiliation{
    \institution{Microsoft Azure}
    \country{United States}
}

\author{\vspace{1pt}Nithish Mahalingam}
\affiliation{
    \institution{Microsoft Azure}
    \country{United States}
}

\author{\vspace{1pt}Brijesh Warrier}
\affiliation{
    \institution{Microsoft Azure}
    \country{United States}
}

\author{\vspace{1pt}Rodrigo Fonseca}
\affiliation{
    \institution{Microsoft Azure}
    \country{United States}
}
\author{\vspace{1pt}Ricardo Bianchini}
\affiliation{
    \institution{Microsoft Azure}
    \country{United States}
}

\renewcommand{\shortauthors}{Lin et al.}



\settopmatter{printfolios=true}

\newcommand{\sysname}[0]{Slasher\xspace}
\newcommand{\microsoft}[0]{Microsoft\xspace}
\newcommand{\azure}[0]{Azure\xspace}
\newcommand{\simulatorname}[0]{Stratosim\xspace}

\begin{abstract}
Datacenters consume many megawatts of power, and regularly encounter scenarios that require modulating their power draw.
These scenarios include datacenter infrastructure failures, power grid failures, grid services, and more, spanning a diverse range of requirements in terms of the power magnitude, the scope of the reduction, the notice time, and other dimensions.
To address these scenarios, we have built \sysname, a general system for modulating the power of \azure datacenters to handle scenarios ranging from individual racks to regional multi-datacenter grid events.
\sysname coordinates datacenter resources with the goal of meeting power targets while minimizing negative impact on hosted workloads.

In this paper, we review the main power modulation scenarios, characterize the power reduction levers using data from production cloud datacenters, describe \sysname 's system architecture, and formulate the cloud datacenter power modulation control problem.
We also develop a high-fidelity datacenter simulator and propose a workload impact model, using them to design and evaluate power control algorithms.
\end{abstract}

\maketitle

\section{Introduction}\label{sec:intro}

Cloud datacenter sites require 10s-1000s of MW of power capacity, contributing a meaningful and rapidly growing fraction of demand on power grids \cite{knittel2025flexible, shehabi20242024, pjm2025forecast,epriUtilitySurvey2024}.
In the existing paradigm, datacenters expect to always consume as much power as they need, up to a limit defined in the contract with the grid operator.
This paradigm is showing signs of a shift, affected by the trends of increased demand for power-hungry AI applications, and the scarce power capacity in the grids where datacenters are being deployed.
At a time of rapid growth desired by cloud providers and their customers, new datacenter deployments are being de-prioritized \cite{ft2024ireland} and postponed \cite{dcd2025gridupgrade}.
In power-constrained regions, new requirements are emerging for datacenters to dynamically reduce power to support the grid~\cite{msft2023quebec,eirgrid21DCconnect,pjmBoardLargeLoadArrangementPlan, EU2026flexibility, EU2026flexibilityprovisions}; power flexibility is expected to unlock 10's of GW of capacity within existing infrastructure \cite{norris2025rethinking}.

Datacenter power flexibility - the ability to change power draw on demand - is an increasingly important function that involves many scenarios, levers, triggers, goal states, and impacts.
Flexibility scenarios can have external triggers, such as grid failures and demand response programs \cite{msft2023quebec, blackout2019lfdd, pgeBIP}, or internal triggers, such as maintenance and failures of cooling/power equipment \cite{microsoft2023electricalfailure, google2022cooling}.
Each scenario can vary significantly in the magnitude of power reduction, how often it occurs, the reduction domain, the event duration, and the amount of advance notice time.
For example, a cooling failure may require a 10\% power reduction in an entire data hall for 3-6 hours while equipment is repaired, with 30 minutes of thermal inertia lead time; in contrast, a concurrent fault across multiple rows may require a near-instant reduction of 5\% power across a subset of affected datacenter power domains to prevent cascading failure.

In a public cloud datacenter, power is dictated by many thousands of servers, each typically hosting one or more virtual machines (VMs), each of which might run one of many thousands of applications, into which the cloud platform has limited visibility.
This limits cloud datacenter operators from broadly adopting power management policies that leverage application flexibility to reduce power, a topic that has been extensively explored in the literature\cite{souza2023ecovisor, lechowicz2025carbon,lin2012dynamic,stojkovic2025dynamollm}.
Instead, operators must rely on data from the platform and power devices,  such as server occupancy, resource utilization telemetry from servers, and power telemetry from meters across the power delivery system.

A multitude of software- and hardware-based levers for power modulation exist across hierarchies of power and IT components within datacenters, including power devices such as generators and batteries; hardware control levers such as server throttling and shut-down; platform levers that manage the state of VMs; and application levers such as workload migration and delay.

Although these levers can effectively shed power in most scenarios when applied individually, each lever has notable financial, performance, and utilization limitations when employed in isolation.
Further, these levers differ in how much power they can reduce,
for how long, and with what advanced notice they can operate.
For example, batteries are expensive when provisioning enough to handle hours of datacenter load, and the use of fossil-fuel (typically diesel) generators is regulated and sometimes prohibited from being used for prolonged periods \cite{uk-generators}.
Levers that involve VMs and servers can impact application metrics and risk violating Service Level Objectives (SLOs) \cite{zhang2021scheduling,xing2023carbon}.
Overall, cloud datacenters present a large and complex action space for power modulation, where the primary challenge is to satisfy a wide range of scenarios while minimizing the negative impact on hosted workloads.

To manage power and support emerging scenarios, we have built \sysname: a general system for datacenter power flexibility.
\sysname manages the power consumption of \microsoft \azure datacenter infrastructure through a combination of power devices and software platform levers, and explicitly minimizes impact to hosted services while meeting power targets.
For each supported scenario, \sysname uses a modulation policy to compute a playbook: a table mapping levers to specific power domains.
When a scenario is triggered, \sysname activates a playbook to modulate power to within the scenario's power target.
Playbooks are periodically recomputed to adjust to the dynamic state of the datacenter.

Computing power shedding playbooks involves dealing with a large optimization space of power reduction actions, inter-dependencies, and power implications.
To experiment with control policies for \sysname, we have built \simulatorname, a datacenter simulator to replay datacenter power telemetry at the granularity of VMs. \simulatorname enables testing of policies that deploy any combination of power shedding levers.
To study the power modulation control problem, we analyze a variety of cloud server and service metrics, and propose a \emph{general impact model} based on capacity and load to estimate the risk of SLO violations caused by power shedding actions for common cloud applications.
We express impact at the scale of a \emph{service}: a set of one or more VMs that compose a cohesive element of an application.
The impact functions are designed to be a faithful proxy to the risk and magnitude of SLO violation caused by power reduction levers, and leverage the (limited) visibility into applications, such as grouped VM deployments, VM size, and resource utilization telemetry.

The development of \sysname is ongoing and evolving. The
core system and mechanism are operational and in production, but some levers and scenarios described in this paper are still being integrated.
In addition, GPU- and LLM-specific workload
management is ongoing work and not discussed in this paper.
Although there are opportunities for power modulation~\cite{patel2024splitwise, patel2024characterizing, liang2026gpu, stojkovic2025tapas, radovanovic2021carbon} when these workloads are present, these techniques require application visibility and control that is generally not available at the platform layer.
Such application-specific power modulation would require direct interaction between platform and workload.
This paper focuses on addressing the challenges of managing public cloud datacenters at the platform layer, leaving platform-workload interaction as future work.

Our contributions are as follows:
\begin{itemize}
\item Analysis of the scope of power reduction from production traces at \microsoft;
\item Design of \sysname, a general system for modulating datacenter power;
\item Results of a live experiment of coordinated battery and workload management for power reduction
\item \simulatorname, a high-fidelity DC power simulator\footnote{\simulatorname will be open-source with the publication of this paper};
\item A general workload impact model, and formulation of the impact-minimization power control problem.
\item Simulation-based experiments of several \sysname control policies.
\end{itemize}

\section{Background}
\label{sec:background}

\subsection{Generalized Power Management Scenarios}
Scenarios that require datacenters to actively manage power include the following broad categories:
\begin{itemize}
    \item \textbf{Power oversubscription mitigation}, where a correlated surge in power across many servers in a power domain must be capped to prevent overloading power infrastructure.
    \item \textbf{Datacenter infrastructure outage}, such as power or cooling system failures that require reducing power to match the remaining capacity.
    \item \textbf{Grid shortages}, where a grid has a long-term energy scarcity.
    \item \textbf{Grid emergencies}, such as a sudden loss of generation capacity that triggers power recovery mechanisms to avoid a cascading blackout.
    \item \textbf{Grid services}, such as peak-shaving programs that enable a grid to oversubscribe peak capacity, or frequency response for maintaining power quality.
    \item \textbf{Grid-interconnection requirements}, such as stabilizing multi-megawatt power fluctuations of synchronized AI workloads and providing low-voltage ride-through during brief grid-voltage aberrations.
\end{itemize}

\begin{figure}
    \centering
    \includegraphics[width=0.85\linewidth]{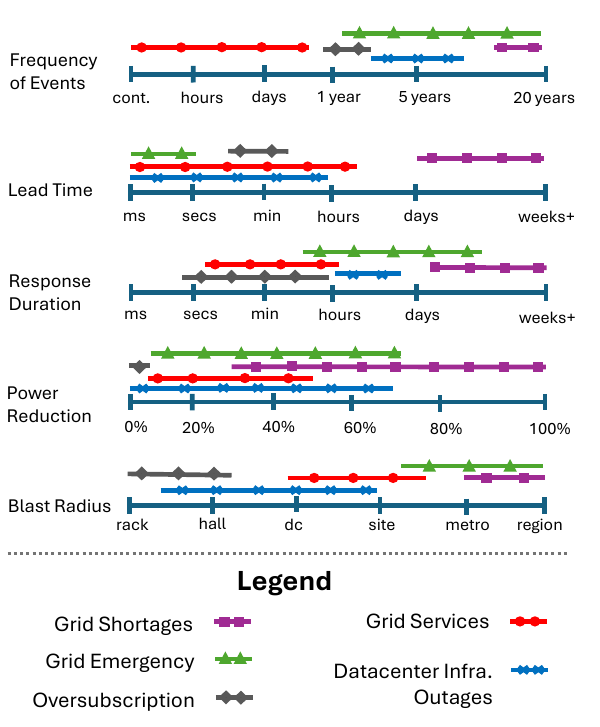}
    \caption{Several broad classes of power modulation scenarios and their key characteristics.}
    \label{fig:dimensions}
    \vspace{-0.1in}
\end{figure}

Grid-interconnection requirements currently demand responses over very short, often sub-second, time horizons and lead times that are unsuitable for software platform-level controls.
We therefore consider these scenarios outside the scope of \sysname.
The remaining scenarios require power reduction\footnote{Increasing power is also within the scope of \sysname, for example by launching dummy workloads; we focus on the more difficult power-reduction scenarios.} with significant variance in their requirements.
We identify five key dimensions that form a \emph{unifying abstraction} for scenarios: event frequency, lead time, response duration, power reduction, and blast radius.
Figure~\ref{fig:dimensions} shows a representative range for each scenario across these dimensions.
The driving vision for \sysname is a single system that handles scenarios across the full range of each dimension.

\textbf{Power oversubscription.}
Existing datacenter power management systems enable power oversubscription while guaranteeing high-availability for a large fraction of workloads while mitigating the effects of power spikes.
When power draw exceeds the capacity of a power device, protective breakers are engaged and cause servers within the affected power domain lose power.
To mitigate this, datacenters use power management systems to cap power and prevent excursions, using workload characteristics, priority, and availability requirements across the power distribution hierarchy to make decisions that minimize workload impact; examples include Meta~\cite{wu2016dynamo}, Microsoft ~\cite{kumbhare2021prediction, zhang2021flex}, and others~\cite{li2019scalable, li2020thunderbolt}.
The main power capping approach is to reduce CPU/GPU frequency, which in theory could be used to support other scenarios.
However, the systems are specialized for timely response to minor power excursions required for oversubscription scenarios and do not inherently support the variety of scenarios that require power modulation (e.g., infrastructure outages or external grid events).
Rather, such systems are subsumed by \sysname, which considers mitigation for oversubscription as one of many supported scenarios.

\textbf{Datacenter infrastructure outage.}
Infrastructure outages and power oversubscription scenarios have the same underlying cause: power demand exceeding the capacity of datacenter infrastructure.
Although individual pieces of infrastructure rarely fail~\cite{uptime2021survey, uptime2025analysis}, they are a regular occurrence at the scale of major cloud providers, and expected to become more prevalent with the rapid growth of cloud capacity, especially in regions with lower power quality. For example, a utility voltage sag may cause failures in equipment that is not protected by a UPS. Cooling infrastructure such as chillers and air handlers can fall under this category; these systems can generally be restarted in response to a power quality event, and these restarts can fail and require minor repair. Loss of cooling capacity may require shedding power to within a lowered cooling capacity budget \cite{bbc2022heatwave, msft2023singaporecooling}. Similarly, failure or maintenance of power infrastructure such as UPSs, transformers or switchboards (often coupled with power oversubscription) may need immediate power shedding across several power domains to prevent cascading failures \cite{zhang2021flex}. These are often local events with varying lead times and long durations while repairs are executed.

\textbf{Grid Shortages.}
Resource shortages that affect the grid's ability to deliver energy long-term can have many causes, including natural disasters, civil unrest, or global conflict.
A recent example is the war in Ukraine and subsequent energy crisis \cite{WEF2022ukraine} that threatened energy scarcity in parts of western Europe.
The defining characteristics of these scenarios is a long lead times and the potential for multi-week durations and large power reduction targets across entire regions.

\textbf{Grid Emergencies.}
Grid failure is region-wide power outage. Although power grids are designed with high reliability standard to avoid such events (e.g. 1 event every 10 years \cite{ercotOutageStandard} or $\le$8 hours/year \cite{eirgrid23report}), datacenter are deployed at a scale where this is likely to affect the cloud platform.

Grid operators are putting plans in place to recover power in the event of large failures by relying on large power consumers, including datacenters.
For example, the National grid in the UK has a Low Frequency Demand Disconnect (LFDD) plan in place in the event of large drops in grid frequency.
Under the LFDD, large customers must either rapidly shed power, else substation breakers trip and remove a fraction of their capacity, between 5-60\% depending on the severity of the frequency drop, within 200 ms \cite{blackout2025lfdd}.

\textbf{Grid Services.}
Grid demand response requires customers to shift or shed power demand following the grid signals \cite{ieaDemandResponseIntro}. As grid power supply becomes increasingly variable due to increased wind and solar generation, demand response is critical for the grid to balance generation and demand. Specifically, in regions where datacenters account for a large fraction of load growth, demand response is being proposed as a requirement~\cite{eirgrid21DCconnect,pjmBoardLargeLoadArrangementPlan} for acquiring grid capacity. In the face of surging of AI workloads, recent studies show that many Gigawatts of datacenters can be connected to the grid if they can occasionally reduce power \cite{lin2024exploding,norris2025rethinking}.

Grid peak shaving is a demand response program that requires enrolled customers to reduce load during peak times due to high load, low generation, transmission limitation, etc. These events are highly predictable (e.g. along with weather forecast, planned maintenance), enabling long lead time to react. The specific constraints on lead time, single-event duration, and invocation frequency are usually specified in contracts \cite{pgeBIP}.

The participation of datacenters in grid response programs has been studied extensively\cite{liu2013data,le2016joint,zhou2018truthful,zhang2021distributed}. Recently, cloud providers have increased their efforts in load shedding and shifting to help stabilize local grids. Google announced temporary reductions in power demand through temporal and spatial shifting when receiving a notice from a grid operator \cite{googleDCDemandResponse}. Microsoft has made the uninterruptible power supply (UPS) system in datacenters grid-interactive in Ireland, where datacenter construction is limited by power constraints \cite{msGridUPS}; the grid-interactive UPS provides fast frequency response to the grid by discharging/charging the battery to balance a supply-demand mismatch.

\section{Power Modulation Levers}
\label{sec:data}

Every datacenter Watt is useful, but some Watts are more useful than others.
Opportunities for low-impact power reduction are often found where resources act as a buffer:
unallocated servers act as a buffer for spikes in VM demand; low-utilized VMs are often provisioned to absorb spikes in requests; idle CPUs rarely sleep to avoid high latency for infrequent requests.
Reducing the power for some of these buffer resources without significant workload impact can require many carefully-coordinated actions, each shedding a small fraction of overall power.

This section describes the most prominent power reduction levers in cloud datacenters. Where possible\footnote{without revealing sensitive information}, we quantify the opportunity using data collected across 20 datacenters at \microsoft over 9 days.
The data includes servers that host CPU compute VMs, and their power consumption is a significant fraction of the datacenter.
Some of the datacenters include recently-deployed clusters that have not accumulated a stable load, while most have been deployed for one year or more.

\begin{table}[h!]
\centering
\renewcommand{\arraystretch}{1.3}
\begin{tabular}{>{\raggedright\arraybackslash}m{4.6cm}*{5}{>{\centering\arraybackslash}m{0.35cm}}}
\toprule
& \makebox[0pt][r]{\adjustbox{angle=-25,origin=r}{Oversub Mitigation}}
& \makebox[0pt][r]{\adjustbox{angle=-25,origin=r}{DC Infra. Outage}}
& \makebox[0pt][r]{\adjustbox{angle=-25,origin=r}{Grid Emergency}}
& \makebox[0pt][r]{\adjustbox{angle=-25,origin=r}{Grid Shortage}}
& \makebox[0pt][r]{\adjustbox{angle=-25,origin=r}{Grid Services}}\\
\midrule
Turn on generator                       & \no & \no & \sometimes & \yes & \sometimes  \\
Discharge energy storage                & \no & \no & \sometimes & \sometimes & \yes \\
Diverting new deployments               & \sometimes & \yes & \yes & \yes & \sometimes  \\
Shut down unallocated servers           & \yes & \yes & \yes & \yes & \sometimes \\
Throttle servers                        & \yes & \yes & \yes & \no & \no  \\
Live migration - out of blast radius    & \sometimes & \sometimes & \sometimes & \yes & \no  \\
Live migration - consolidation          & \sometimes & \sometimes & \sometimes & \yes & \no  \\
Shut down allocated servers             & \yes & \yes & \sometimes & \no & \no \\
\bottomrule
\end{tabular}
\small Legend: \yes : Yes, \sometimes : Some scenarios, \no : No
\caption{Mapping software-controlled levers to the classes of scenarios where they are appropriate to use.}
\label{tab:actions_scenarios}

\end{table}

Our position is that these workload management levers are best deployed sparingly to support rare emergency events, and do not support scenarios with larger power reduction requirements and/or occur more than once every few years.
Table~\ref{tab:actions_scenarios} summarizes the set of actions that would be reasonable to apply to each scenario class.
In the rest of this section, we break down datacenter power draws into categories, and discuss each levers potential power reduction and impact.

\begin{figure}
    \centering
    \includegraphics[width=0.8\linewidth]{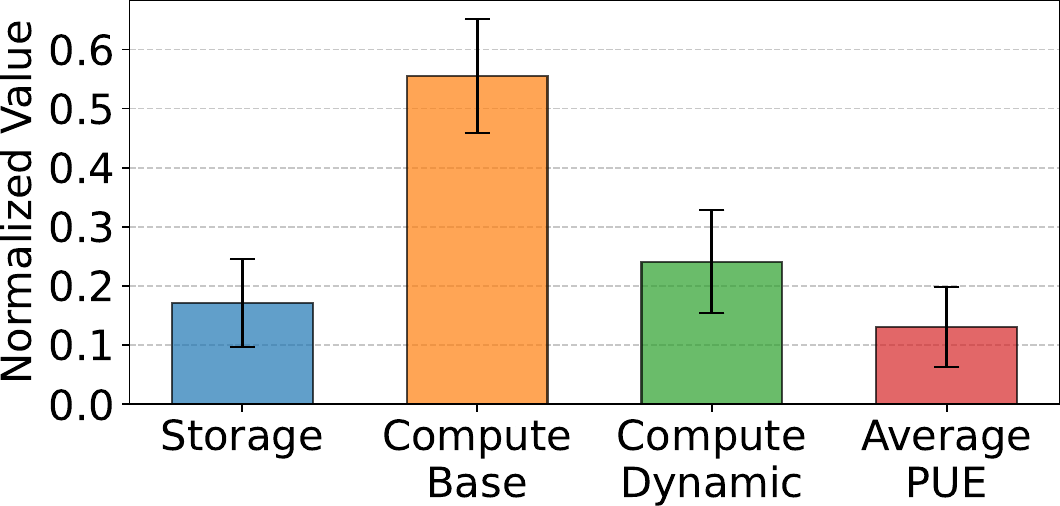}
    \vspace*{-1.0em}
    \caption{Fraction of IT power within datacenters, normalized to maximum observed power usage; PUE is included as fraction of maximum IT power.}
    \label{fig:dc-power-breakdown}

\end{figure}

\subsection{Batteries and Generators}
Many scenarios can be handled with a combination of batteries and generators without affecting the workloads; the caveat is that they can't be used for scenarios that require reducing IT power, such as cooling failures.
Batteries can be excessively expensive when provisioned for multi-hour events. Generators cannot always be deployed due to local regulations and their operating durations are often restricted.
Workload management levers can be viewed as complimentary, to close the gap for handling internal failures and rare grid events to avoid provisioning expensive battery capacity for tail events, especially where generator use is restricted.

\subsection{Power breakdown within datacenters}

Figure~\ref{fig:dc-power-breakdown} presents a breakdown of average datacenter power to identify opportunities for reduction.
It includes storage servers; general-purpose compute server power, divided into base (idle) and dynamic components; and cooling and power-delivery overheads, represented by PUE.
The key observation is that \textbf{the base power of compute servers dominates total datacenter power}; therefore, the largest opportunity for power reduction lies in turning off servers.
Dynamic power, which arises from VM utilization of server resources, can be recovered by turning off VMs or reducing processor frequency.
GPU servers (not shown) exhibit a different power breakdown because their base power constitutes a smaller fraction of total server power than that of CPU servers.
Although not explicitly studied in this paper, datacenter storage services and cooling systems also present meaningful opportunities for power reduction.

\subsection{Shutting down servers}

A large fraction of datacenter power is in the base power consumed by servers; this power can only be shed by turning them off or putting them into low-power states where they remain online but not fit for hosting VMs.
Some fraction of servers in cloud datacenters are unallocated, though the number varies across datacenters with <5\% common in mature clusters and >20\% seen in recent deployments.
Spare unallocated servers are put into low-power states already, and turning those servers off does not reduce power substantially.
The cloud platform provisions additional servers to keep them unallocated and provide a buffer for demand growth and failovers from existing servers; turning them off can substantially increase the risk of VM deployment rejections, which can have a major impact on workloads.
The impact associated with this risk is relatively low for the first several servers in the buffer, but increases rapidly as the buffer count shrinks to 0.

The vast majority of servers are hosting VMs.
Shutting down such servers can have a large impact on the applications running on these VMs and is a last resort, although the relative impact depends on the resilience and capacity provisioned for the affected application.

\subsection{Throttling Processor Frequency}

\begin{figure}
    \centering
    \includegraphics[width=\linewidth]{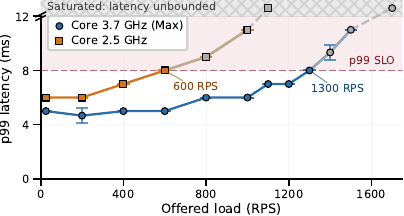}
    \caption{Vault's p99 latency vs.\ offered load, with core-frequency setpoints at maximum frequency (3.7~GHz) and 2.5~GHz; the dashed line is the workload's p99 SLO and points in the gray region indicates saturation when queuing latencies grow unbounded. Labeled points are the serving capacity for each setpoint. Lowering the core frequency reduces serving capacity from 1300 to 600~RPS, i.e., 46\% of its peak load.}
    \label{fig:slo_knee_example}
\end{figure}

\begin{figure}
    \centering
    \includegraphics[width=\linewidth]{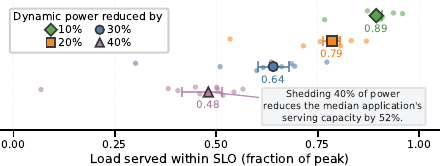}
    \caption{Serving capacity retained within each application's SLO at exact dynamic-power reductions of 10--40\%, normalized to its maximum-frequency serving capacity. Dynamic power is relative to the maximum-frequency setpoint. Dots show the ten applications, markers show the median, and whiskers span the inter-quartile range. At 40\% less dynamic power, the median application retains 48\% of its peak serving capacity. These results suggest the reduction in power from frequency capping is proportionally smaller than the resulting loss in serving capacity.}
    \label{fig:serverpower_freq_throughput}

\end{figure}

Server power consumption and workload throughput are highly sensitive to CPU frequency setpoints.
To characterize this power–throughput tradeoff, we study ten request-serving workloads (described in Appendix~\ref{app:workloads}).
Each workload runs in a single 8-vCPU VM pinned to dedicated cores of a 28-core Intel Xeon Gold 5512U (Emerald Rapids, 2.1~GHz base, 3.7~GHz maximum turbo frequency, SMT enabled), with the rest of the cores idle.
Our experiments isolate the effects of the core frequency: the uncore frequency, which drives the shared cache and memory interconnect, stays at its maximum.
This analysis complements an extensive set of studies of LLM inference on GPUs~\cite{patel2024characterizing, stojkovic2025dynamollm} which show a wide variety of throughput impacts across workload characteristics such as model type and input-output token ratios.

Figure~\ref{fig:slo_knee_example} illustrates workload throughput at a given frequency using profiling results for the secret-store Vault~\cite{hashicorp2026vault} at two setpoints.
Each curve plots p99 request latency against offered load, measured in requests per second (RPS), at one core-frequency setpoint.
Sweeping the offered load at maximum frequency identifies the workload's peak load—the knee of the curve—beyond which queuing delay dominates and latency increases sharply.
Vault reaches this knee at 1300~RPS, with a p99 latency of 8~ms, which is used as the workload's SLO.
A VM's \emph{serving capacity} for a given application and frequency setpoint is therefore the highest RPS it can sustain while meeting that SLO.
At 2.5~GHz, Vault's serving capacity decreases to 600~RPS, or 46\% of its peak load.

The reduction in serving capacity that comes with frequency capping is a worthwhile trade-off only if it results in a comparable reduction in power.
Figure~\ref{fig:serverpower_freq_throughput} shows how much serving capacity remains across all ten applications for dynamic-power reductions from 10\% to 40\%.
For each application and measured core-frequency setpoint, we linearly interpolate between measured points to find the frequency setpoint's serving capacity and its power between adjacent offered loads.
The figure plots serving capacity normalized to the capacity at maximum frequency.
Thus, we characterize how much of an application's peak load a VM can serve within its SLO when it must shed a given fraction of its dynamic power by reducing the core frequency.

Most workloads lose more serving capacity than the achieved power reduction, but the capacity loss varies widely across applications.
At 40\% less dynamic power, retained serving capacity ranges from 7\% to 57\% of peak across the ten applications.
The compute-bound Go HTTP service retains the least serving capacity at every power-reduction target: 77\% at 10\% less dynamic power and 7\% at 40\% less dynamic power.
The median application retains 89\% at 10\% less dynamic power, 79\% at 20\%, 64\% at 30\%, and 48\% at 40\%.
Thus, CPU frequency reduction tends to shed proportionally less power than the serving capacity it removes; in terms of power-per-capacity, turning off the VM can offer a better trade-off, especially for workloads that are horizontally scaled across many VMs.
Frequency throttling is therefore best suited to rare, short-lived scenarios, such as mitigating utilization peaks in the datacenter power delivery network~\cite{li2020thunderbolt, zhang2021flex}.

\begin{figure}
\begin{subfigure}{.5\textwidth}
  \centering
  \includegraphics[width=.9\linewidth]{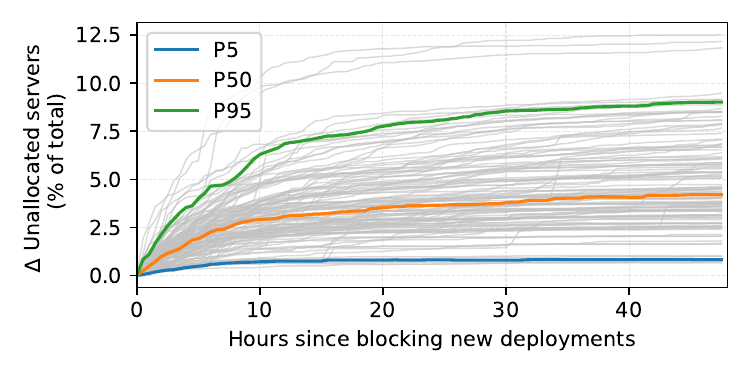}
  \label{fig:unallocated_servers}
\end{subfigure}
\begin{subfigure}{.5\textwidth}
  \centering
  \includegraphics[width=.9\linewidth]{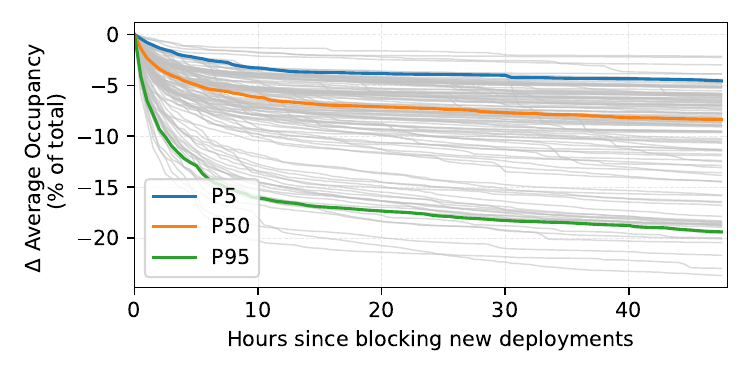}
  \label{fig:average_occupancy}
\end{subfigure}
\caption{Unallocated servers (top) and average server occupancy (bottom), relative to hour 0, after preventing new VM deployments to the datacenter.}
\vspace{-0.1in}
\label{fig:stop_deploy}
\end{figure}

\subsection{Diverting new deployments}
Diverting or reducing new VM deployments to a specific datacenter, or any event blast radius, can reduce power as existing VMs gradually depart and server utilization drops.
This is a low-impact lever, as the vast majority of VM deployments do not have a locality placement constraint and can be routed away from the blast radius to other data halls or datacenters in the region.
Its limits are defined by the available capacity outside of the scope of the event, and cloud platforms generally aim to minimize unallocated capacity; nevertheless, for large regions with 10's of datacenters, some capacity is typically available for temporary diversions.
This action is less useful for region-wide scenarios, which includes many grid services, since datacenters within a region often share a power grid.

Figure~\ref{fig:stop_deploy} shows, upon stopping new deployments to a datacenter, the change in unallocated servers (top) and average server occupancy\footnote{defined as the maximum of allocated CPU cores and allocated memory fractions at the server level} (bottom) metrics.
This data was generated by analyzing VM allocation and de-allocation data at \microsoft and projecting server states if no new VMs are allocated after 00:00 UTC for 7 different days across 20 datacenters, and each trace represents a single datacenter-day combination for a total of 140 examples.
The number of unallocated servers and occupancy changes quickly within 2 hours, with a 2\% increase in unallocated servers and roughly 8\% decrease in occupancy for the median trace, caused by VMs that re-deploy on the hour to run short workloads.
The rate of change in both metrics slows considerably past 12 hours, which represents the departure of many VMs with daily deployment periodicity, at which point 3-7\% additional servers become unallocated and a 5-15\% reduction in average occupancy.
The highest percentiles of these traces correspond to datacenters with many recently-deployed clusters that have not yet established a large population of long-lived VMs; such datacenters can reduce their power meaningfully by blocking new deployments, but the opportunity lasts for a short period in their life cycle.

\subsection{VM migration and consolidation}

\begin{figure}
    \centering
    \includegraphics[width=0.95\linewidth]{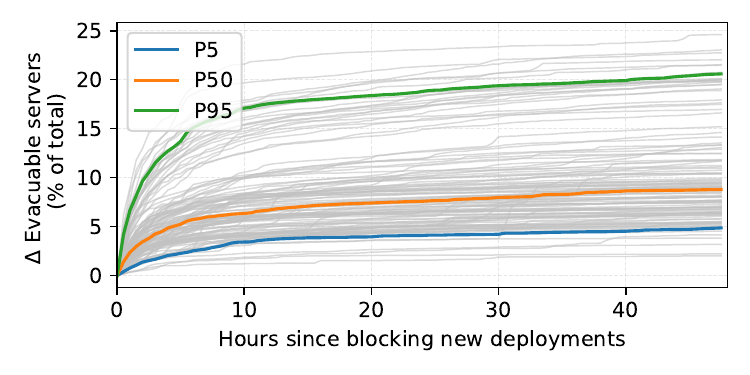}
    \caption{Newly de-allocated servers after consolidation in an ideal case where all VMs are eligible for live migration and new deployments are turned off at hour 0.}
    \label{fig:consolidate}

\end{figure}

Live migration of VMs has multiple uses in power modulation.
VMs could be consolidated onto a smaller set of servers, which allows for de-allocated servers to be put into a low-power state.
Short of full evacuation, some VMs could be migrated off a server before turning it off to mitigate the impact of a shut down.
VMs can also be migrated away from the blast radius of an internal datacenter event, such as a cooling failure within a single data hall.
However, the VM migration process, which involves mirroring the memory on two hosts simultaneously, can contend for the CPU and affect server performance during the transition.
These impacts are short-term and thus relatively minor compared to losing the VM entirely.
The main limitation of migration is its duration, which limits uses to scenarios where the lead time is on the order of 10's of minutes or longer.

Consolidation-via-migration is more effective when combined with prevention of new deployments, which decreases occupancy over time and allows for more consolidation.
Figure~\ref{fig:consolidate} shows the number of servers that can be de-allocated after consolidating existing VMs onto the minimum set of servers, with a 10\% capacity buffer left unused on each server that does not initially exceed 90\%, to approximate stranded capacity.
Additional servers can be de-allocated after turning off new VM deployments, with a median of 3\% additional servers after one hour, and 8\% after 24 hours.
These numbers are idealistic, since several practical factors can limit the scope of consolidation, including non-deterministic VM migration times and VM features that prevent migration; notably, a single migration-ineligible VM prevents a server from being evacuated.

\section{\sysname System Design} \label{sec:systemdesign}

\sysname is designed to manage datacenter power by  jointly controlling the power infrastructure (generators and batteries), the IT infrastructure (servers), and VMs. In this section, we make the case for a centralized power management system, describe the system requirements, present the system architecture, and show an example of hardware and software levers in action in a live datacenter for a Fast-Frequency Response (FFR) grid service scenario.

\subsection{Centralized hierarchical design}

Centralized, region-level control is motivated by three factors: 1) coordinated response to events that include multiple datacenters, 2) the opportunities and risks associated with managing simultaneous scenarios that overlap in scope, and 3) the engineering benefits of maintaining a single system.

\paragraph{Multi-datacenter coordination}
Grid-level scenarios can have multiple datacenters within their scope.
Rather than having a utility company decide how much power each datacenter should shed, a centralized controller can reason about the amount of impact a reduction of power would have at each datacenter.
For example, datacenters that are primarily hosting many high-priority workloads can be avoided, while those with lower-priority workloads can be prioritized.

\paragraph{Simultaneous scenarios} Managing multiple simultaneous scenarios requires coordination that can be difficult to manage across many systems.
For example, consider a case where a row of racks is experiencing a fault and requires power reduction, while at the same time a rack in that row is approaching its power limit.
A centralized system can recognize the opportunity to manage both scenarios with a single action, such as turning off VMs on the power-constrained rack or migrating them away from the row. Individual scenario handlers may miss such opportunities, or make the problem worse by initiating actions that help one scenario at the expense of another.

\paragraph{Engineering benefits} As a point of contrast, consider a separate system built to manage each scenario; in this case, updates to hardware or software lever interfaces requires updates to every system that uses them, and new scenarios would require building and maintaining new systems.
The unified abstraction described in Section~\ref{sec:background} lends itself to a single system built to support the full range of power modulation scenarios.

The challenge with centralized design is system scale, which manifests in both the large action space—including individual VMs and servers—and the volume of data informing power-modulation decisions.
In \sysname, this scale is managed through a hierarchical design in which a centralized region-level controller configures and receives data summarizes from local per-datacenter controllers.

\iftrue
\begin{figure}[ht]
  \centering
  \begin{subcaptionbox}{Regional Orchestrator\label{fig:arch:orchestrator}}[\columnwidth]
    {\includegraphics[width=0.75\columnwidth]{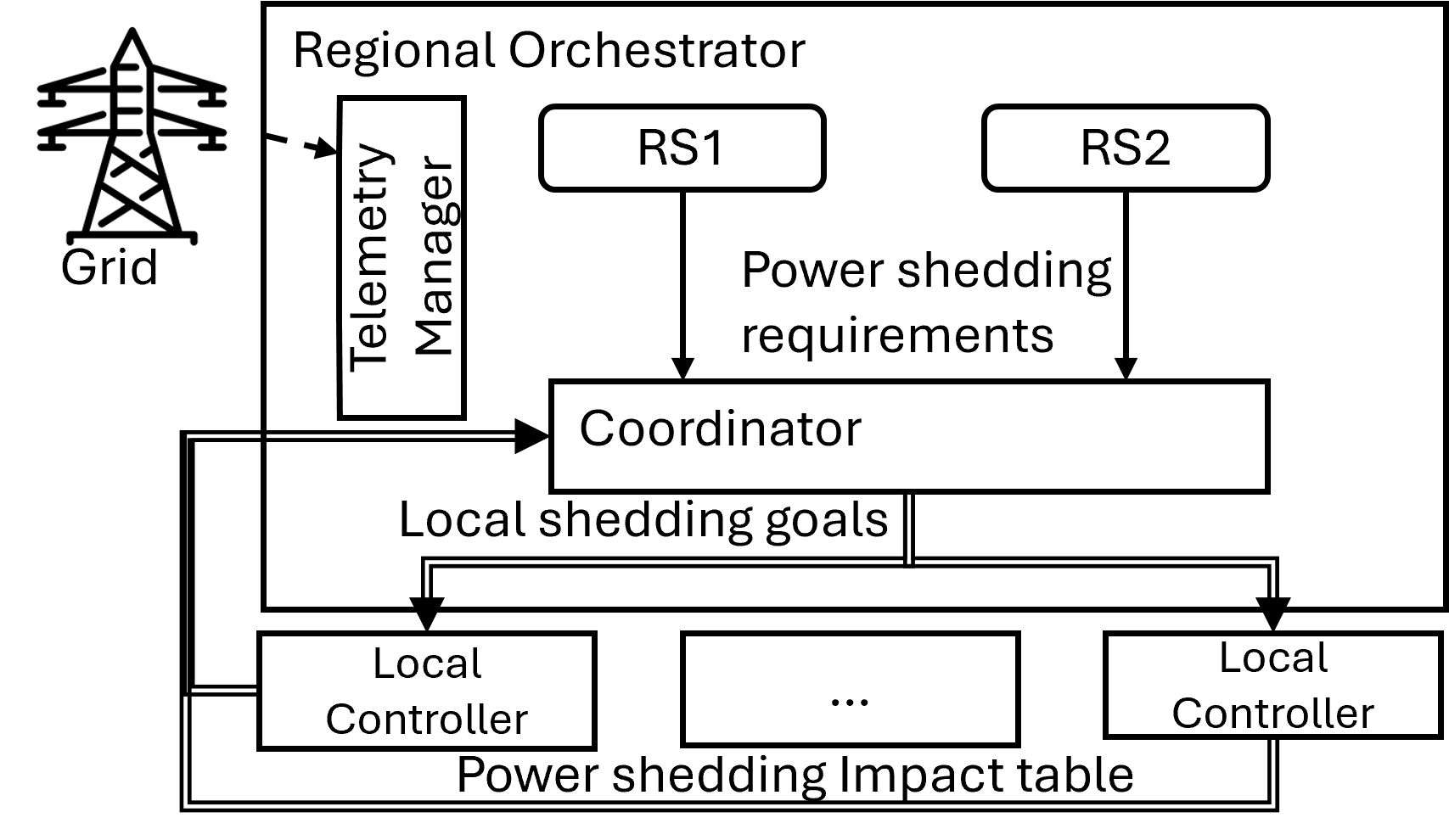}}
  \end{subcaptionbox}
  \hfill
  \begin{subcaptionbox}{Local controllers\label{fig:arch:controller}}[\columnwidth]
    {\centering
    \includegraphics[width=0.9\columnwidth]{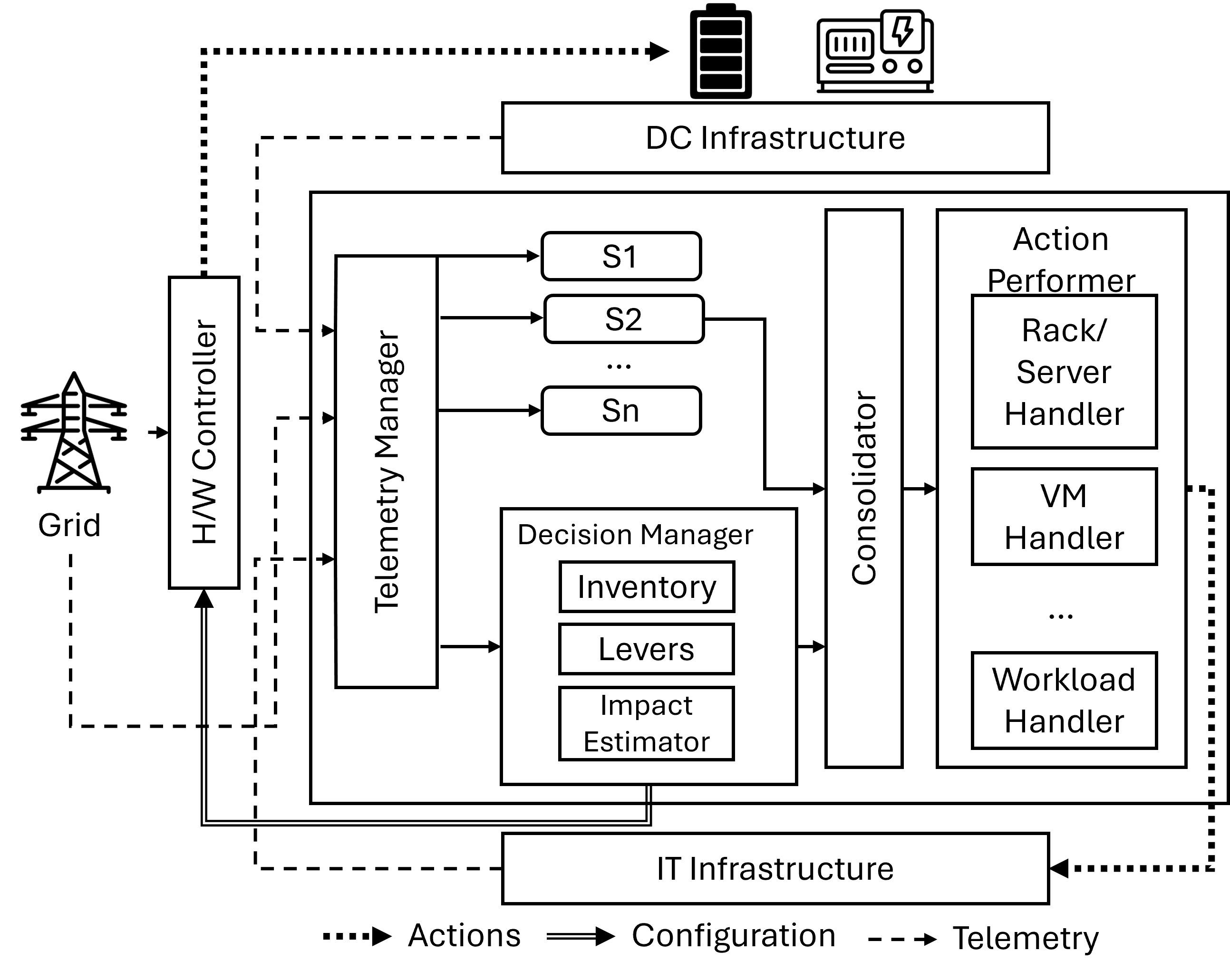}}
  \end{subcaptionbox}
  \caption{Slasher: Architecture Overview}
  \label{fig:arch:overview}

\end{figure}
\fi

\subsection{Requirements}

\sysname is designed to operate across the five scenario dimensions presented in \figurename~\ref{fig:dimensions}.
First, blast radius: it must handle both hall-local disturbances (e.g., a cooling loop failure) and regional events that span multiple datacenters (e.g., grid demand response). Second, lead time: it must react to scheduled, day-ahead dispatch, seconds-scale infrastructure faults, and millisecond-scale frequency deviation events. Third, response duration: it must sustain minutes-long trims, peak-hours shedding, and multi-day curtailments. Fourth, shedding magnitude: it must deliver everything from small 1\% nudges to complete load drops as per the requirements of the scenarios. These varying requirements are reflected in the design, which is hierarchical with pre-armed fast paths, staged and reversible actions, and explicit impact modeling to prioritize user experience and "SLAs."

\subsection{Architecture overview}
To address the blast radius and lead-time requirements, a key design element is the separation of the system into two tiers: \textit{Local Controllers} (Fig. \ref{fig:arch:orchestrator}) and the \textit{Regional Orchestrator} (Fig. \ref{fig:arch:controller}). Local controllers run in each data hall, own high-fidelity telemetry and actuation of power shedding levers, and resolve concurrent local scenario activations. A regional orchestrator supervises regional or cross-data-hall events: it ingests activation signals (e.g. regional grid services), consolidates site-level impact information, and assigns power shedding goals to a subset of halls that collectively meet the regional target with minimal impact through the local controllers. This split localizes detailed resource knowledge to allow for fast, local detection and actions, while enabling globally optimal allocation and blast-radius containment.

\paragraph{Local controller}
The role of the local controller is to monitor telemetry and signals, and activate corresponding levers to shed required power. A key requirement is the ability to respond within short lead-times. It is thus important to pre-compute the responses since gathering required telemetry and making decisions on critical path may prevent a timely response. The local controller (Fig. \ref{fig:arch:controller}) operates in two modes: i) a non-critical or \textit{offline} mode, which periodically aggregates information about the data hall (e.g. power consumption, workload priorities, shed-able power etc.) and calculates in advance what set of actions need to be taken in case of a scenario activation, and ii) a critical or \textit{online} mode that handle scenario triggers and activate levers, performing the power shedding actions within the given time limits.

In the offline mode, a \textit{Telemetry Manager} continuously ingests telemetry from grid, the datacenter infrastructure (power, voltage, frequency, thermal headroom) and IT (server power, utilization, VM priorities) that feed both the \textit{scenarios} and the \textit{Decision Manager}. The Decision Manager maintains the current inventory (racks, servers, VM placement, batteries/generators) and a catalog of levers (e.g., cap or shut down racks/servers, pause or consolidate VMs, throttle workloads, discharge batteries, start generators). Based on received telemetry, lever impacts and real-time priorities, an \textit{Impact Estimator} synthesizes two artifacts for the critical path: a \textit{Decision Table} or a \textit{playbook},  mapping incremental power shed to cumulative actions to be taken, and an \textit{Impact Table} that maps power shed to the corresponding impact of those actions. This allows \sysname to select a subset of actions to achieve the target power goal while minimizing the total impact (for example, as defined in Sec. \ref{sec:impact}).

In the online mode, each scenario ($S_n$) monitors relevant signals and is responsible for activating or deactivating itself, and proposing a data-hall power limit. In cases where scenarios are responsible for aspects other than power (e.g. cooling headroom), it is the responsibility of the scenario to convert them into the corresponding power limits.

In case multiple scenarios are activated at the same time, the \textit{Consolidator} arbitrates the actions by enforcing the most restrictive limit, then looks up the minimal set of actions in the decision table sufficient to meet that limit. The \textit{Action Performer} executes the chosen actions via specialized handlers for racks/servers, VMs, and workloads, with ordering, retries, rate limits, and reversibility to avoid power oscillations and undue disruption.

Further, \sysname can handle scenarios requiring sub-second response times, for which the centralized online path is too slow.
Dedicated \textit{Hardware Controllers} operate as sidecars to \sysname and are physically located near the hardware, such as on-rack control boards or battery management systems.
These devices monitor high-frequency electrical signals and rapidly trigger actions such as battery discharge or processor throttling.
Through the \textit{offline} path, \sysname's decision manager periodically sends scenario-specific configurations to each hardware controller, including thresholds, charging and discharging rates, durations, and server priorities.
As storage devices approach their energy limits or longer responses become necessary, the local controller supplements these fast actions with slower power-shedding mechanisms.

The fast- and slow-paths, coupled with the hardware and software controllers allows the system to address varying lead-times, response durations and magnitudes by combining the fast response through energy storage devices and long-duration, sustained response of software actions.

\paragraph{Regional orchestrator}
The \textit{Orchestrator} (Fig. \ref{fig:arch:orchestrator}) manages regional scenarios (RS1, RS2) that require coordinated response across local per-datacenter controllers. Local controllers periodically publish their impact tables to the orchestrator. The \textit{Coordinator} aggregates these tables and solves a budget allocation problem: it selects a subset of halls and assigns per-hall power shedding goals that achieve the regional target while minimizing the sum of predicted impacts. It then disseminates these goals with ramp profiles and other requirements to individual local controllers. Both these communications occur periodically, ensuring that the most recent impact tables and corresponding goals are available for making real-time decisions on the critical path when an event occurs. This event signal may be monitored locally by the controllers where feasible or communicated by the orchestrator on-demand for scenario activations.

This two-tier design satisfies the \sysname operations requirements across the four axes, \textit{i.e.,} blast-radius, lead time, duration, and magnitude: local controllers deliver precise, fast and sustained actuation with explicit impact minimization, while the orchestrator allocates regional goals.

\subsection{Scenario tracking}

\begin{figure}
    \centering
    \includegraphics[width=0.99\linewidth]{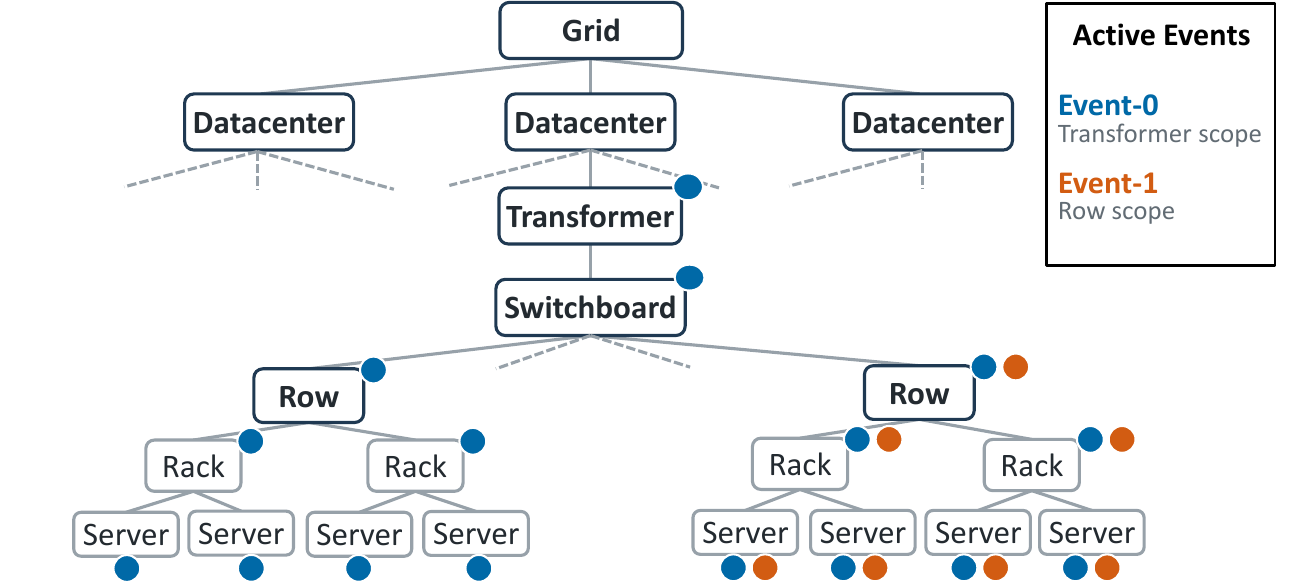}
    \caption{Tree for tracking events across power delivery systems, shown with two simultaneous events.}
    \label{fig:power_event_tree}
\end{figure}

\sysname tracks scenarios by maintaining a tree structure that represents the active power paths in the power distribution centers of every datacenter in the region.
Figure~\ref{fig:power_event_tree} illustrates an example with concurrent events: Event-0 is a transformer-level event, and Event-1 a row-level event. Leaves in the tree represent individual servers, and intermediate nodes correspond to power devices in the distribution system.
When a scenario is triggered, the scenario's scope is mapped to a node in the tree, and all downstream nodes are marked with a unique event id to indicate an upstream event.
For example, a grid-level event would mark every node in the power paths of the affected datacenters, while a row-level event would only mark that row and the racks and servers connected to it.
The presence of multiple marks indicates multiple concurrent events with overlapping scope.
When the event ends, the corresponding markings are removed from the nodes.

The decision engine can reason about simultaneous events by tracking the state of the tree, and account for the effects of an action on all active events within the action's scope when making decisions.
Actions on resources within an overlap can then be prioritized by weighing the combined effect on power reduction across the relevant scenarios.
For example, in the case Event-0 and Event-1 in Figure~\ref{fig:power_event_tree}, turning off a server within the overlapping row counts toward both events, while turning off an identical server on another row only counts toward Event-0.

\subsection{Robustness}
\sysname local controllers process several data streams covering power, utilization, and other relevant metadata for estimating workload impact of actuating over each VM.
Data quality is ensured by existing processes that are not specific to \sysname, such as redundant meters and cross-validation of data with upstream meters in power delivery systems.
Data freshness SLOs are on the order of a few seconds, although gaps can occur when telemetry agents or streams fail; such problems are handled with best-effort approaches, such as filling short gaps with the most recent observations, and removing the affected scope from consideration for longer data outages.

On responding to a scenario, the power impact is continuously monitored and actions adjusted to account for power deficits or surpluses outside of a small tolerance band that is scenario specific.
For example, for oversubscription, power devices are typically protected by breakers that remain closed for several minutes if overdraw is relatively small (eg. <10\%), which provides tolerance for brief spikes.
In contrast, the failure of a large power device such as a transformer would cause a sudden increase in power draw across remaining devices in a distributed-redundant distribution scheme, and the tolerance on power draw is generally within the actual breaker limit to avoid any overage.

\subsection{Power modulation in action} \label{sec:pilot}

\begin{figure*}[ht]
  \centering
  \includegraphics[width=\linewidth]{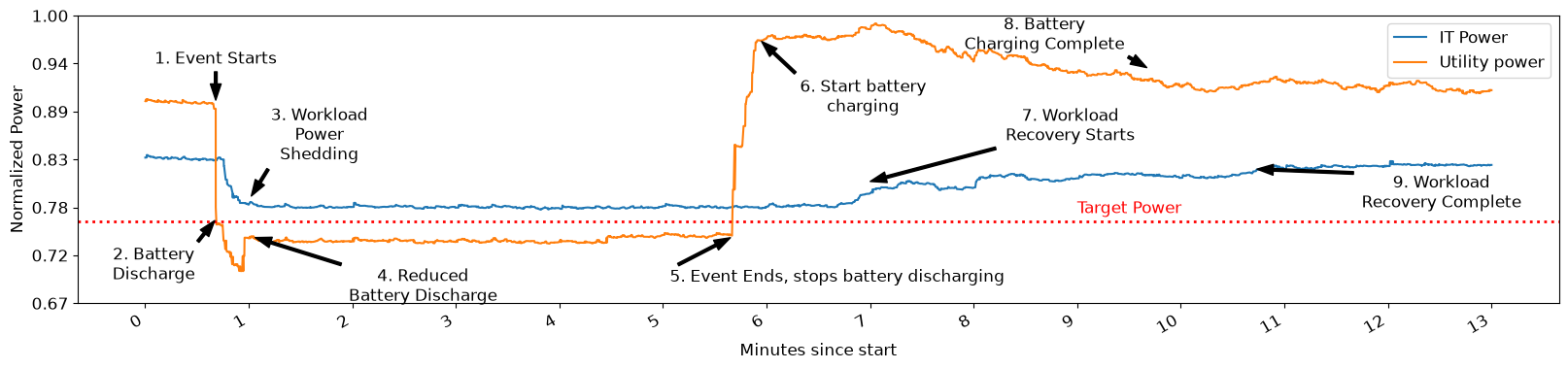}
  \caption{Telemetry collected during a power event with a combined response across power and IT infrastructure levers.}
  \label{fig:flexgups}
\end{figure*}

\sysname is currently deployed across \azure to manage a variety of power modulation scenarios using server shutdown lever, with additional levers in progress. As part of the development process, we ran a pilot study to demonstrates the feasibility of \sysname using a combination of energy devices and IT power controls, which represents the next step in datacenter power management.
THe pilot ran in a live datacenter with hardware-controller, software-configurable UPS, and an IT controller.
The scenario mimics a prolonged fast-frequency response (FFR) event, which requires a sub-second response to be sustained for several minutes.

Figure~\ref{fig:flexgups} shows a timeline of events for the pilot study. In the offline path, the decision manager calculates the amount of power to be shed for a certain drop in frequency, selects the combination of battery discharge and power to shed by shutting down a subset of IT racks, and configures the hardware controller to discharge batteries at a certain rate.
When an offending frequency is observed, the scenario is activated ($\#1$), at which time the hardware controller detects and immediately starts discharging the battery ($\#2$) to reduce power below the power shedding target. The IT controller initiates shutdown of racks hosting \microsoft internal workloads to reduce IT power - with minimal impact - ($\#3$) that activate after a brief delay in the software telemetry pipeline, dissemination, and implementation of selected actions.
When IT power is reduced, the decision manager signals the hardware controller to reduce the battery discharge rate ($\#4$), increasing utility power while staying below the required limit.
As the frequency of power grid restores, the event ends ($\#5$), the battery starts charging ($\#6$) at a slow rate, and the workloads are restored ($\#7$). Once the battery recharge is complete ($\#8$), the workload can also be fully restored ($\#9$) by the Action Performer, and the system goes back to normal state.

\section{Datacenter Simulation} \label{sec:simulation}

\sysname decisions impact both cloud workloads and the power grid, and thus need to be thoroughly studied in a de-risked environment.
To facilitate the study of control policies, we built \simulatorname, a datacenter simulator that replays datacenter state using telemetry streams and model the effect of \sysname actions across the power delivery hierarchy.

\begin{figure}[h]
    \centering
    \includegraphics[width=0.8\columnwidth]{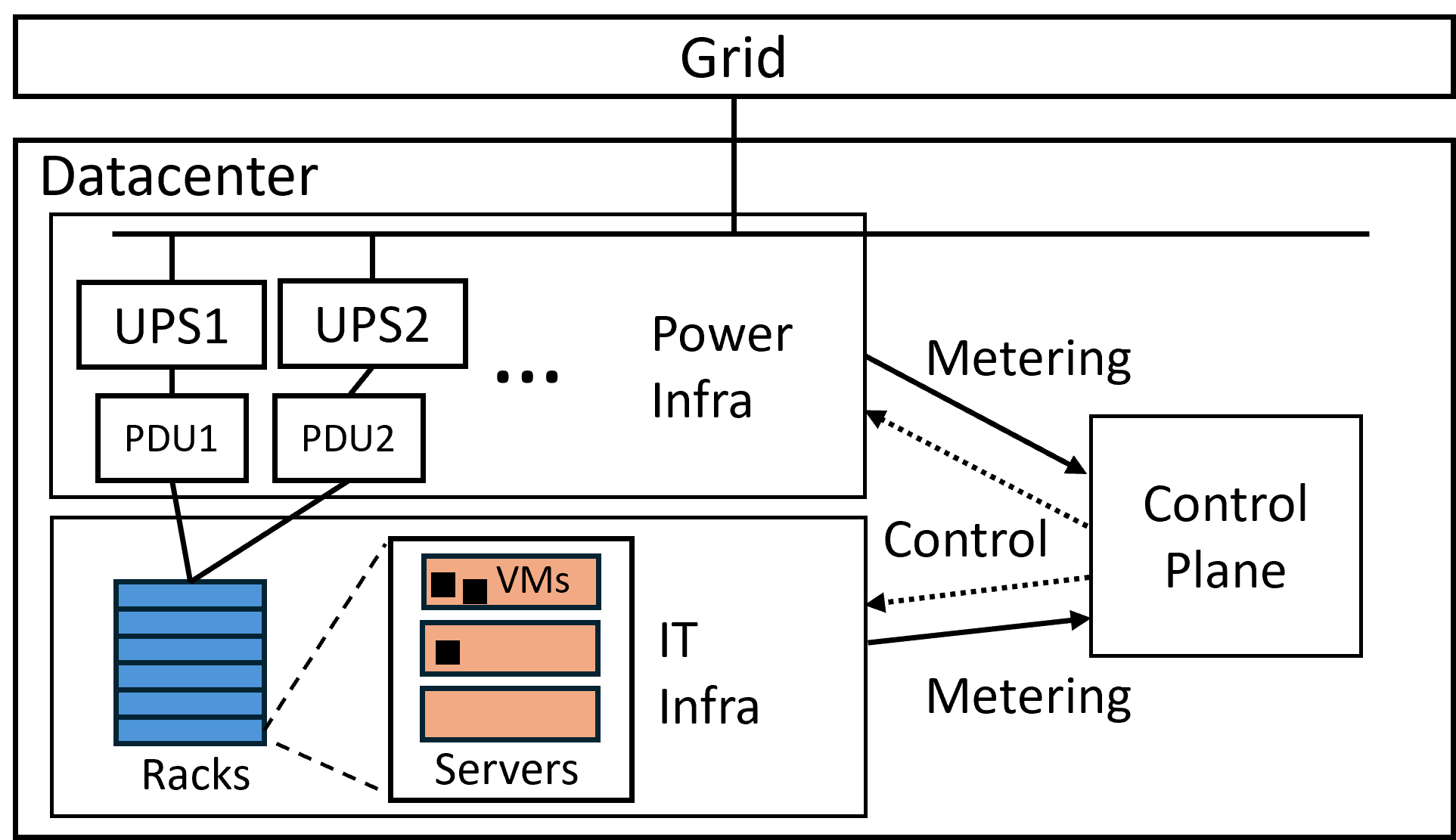}
    \caption{Overview of simulated datacenter infrastructure.}
    \label{fig:DCInfraOverview}
\end{figure}

\paragraph{Simulated Infrastructure}
Figure \ref{fig:DCInfraOverview} shows a reference infrastructure of a cloud datacenter. The UPS works as power conditioner in normal operation and backup power in case of grid outages. Power is supplied to the racks through power distribution units (PDUs), and each rack is connected to two PDUs for redundancy. The specific power delivery network can be configured to model arbitrary topologies.

Servers are arranged in racks and each server can host multiple VMs.
The software controllers in the control plane monitor the metering data from different levels of the hierarchy and adopt control actions (e.g. power capping, shut down servers) when necessary.

\paragraph{Simulator Design and Implementation}

\begin{figure}[h]
    \centering
    \includegraphics[width=\columnwidth]{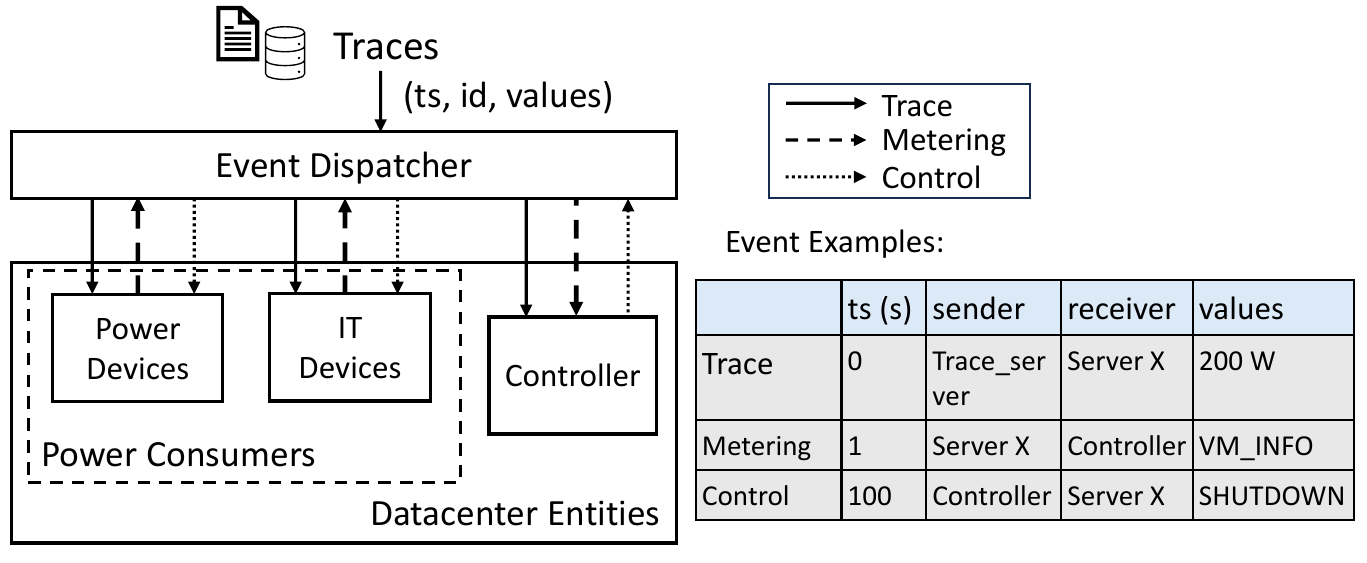}
    \caption{Overview of simulator design.}
    \label{fig:simulatorOverview}
\end{figure}

The simulator is implemented with Salabim, a Python package for discrete event simulation \cite{van2018salabim}. Figure \ref{fig:simulatorOverview} shows an overview of the design and examples of events. Traces of events, including power and resource usage telemetry as well as general event occurrences, are read as (timestamp, receiver id, values) tuples and sent to the receivers by the event dispatcher. The event receivers/senders are the different entities in the power and IT infrastructure shown in Figure \ref{fig:DCInfraOverview}.
Power consumers include the power-consuming IT hardware and upstream electrical hardware. The power consumption is aggregated from the VMs/servers/racks to the grid, with metering events generated. The controller monitors the telemetry and event occurrence. If any pre-defined events happen, it sends control actions to the involved hardware as events through the dispatcher. Key features of the simulator include:
\begin{itemize}
    \item \textbf{Support for diverse trace data sources}: traces can be read from databases, files etc. using connectors or generated at runtime.
    \item \textbf{Large scale and full coverage of the stack}: model the main power and IT infrastructure, from the power grid connection to VMs.
    \item \textbf{Mirroring the reality}: the connections between the real-world components are mapped to the simulator objects, and realistic elements such as communication delays in telemetry and control are represented.
\end{itemize}
This simulator provides a high-fidelity simulation and supports the study of a wide variety of datacenter power scenarios and to evaluate controller policies, which are described in the Sections~\ref{sec:controlpolicies} and~\ref{sec:caseStudy}.

\section{Workload Impact Formulation}
\label{sec:impact}
A cloud datacenter hosts a set of VMs on a set of physical servers.
VMs are typically not unique; we refer to an \emph{application} as a set of VMs that work together to deliver units of work, e.g., user requests.
When deciding which VMs/servers to throttle or turn off, it is important to consider the structure of the applications they compose.
In this section, we describe a set of application archetypes based on characteristics related to the impact of capacity loss, and present a general model for the application impact of power modulation.

\subsection{Service Model}
An application is composed of one or more VMs, and can be simple monoliths or complex microservices with request pipelines.
Within an application, VMs are typically organized into one or more \emph{services}, each executing a separate logical function for the application.
Services have several first-order defining features, including:
\begin{itemize}
    \item User-facing vs. background tasks.
    \item Stateful vs. stateless services, referring to data dependencies between service requests and specific VMs.
    \item Resilience, in particular to regional, zonal, local (within the datacenter) loss of capacity.
\end{itemize}

These three dimensions offer insight into the potential impact of forced shutdown or throttling of a VM.
For example, VMs running user-facing workloads may be more sensitive to latency SLOs under throttling than those running background/maintenance tasks.
For a stateful service, ongoing user sessions would be lost if the VM is turned off, and any requests requiring local data would no longer be serviceable if all corresponding VMs are shut down; a stateless service does not lose critical data when a VM is lost, and requests could be propagated to remaining service VMs.
Resilience has a similar effect, with services designed to remain online for the given failure domain, at the cost of resource provisioning and data replication.

Each service typically has slack capacity: computational capacity in existing VM deployments that is not always used.
A common source of slack is static capacity provisioning for peak load, but even autoscaled deployments provision sufficient capacity to handle short-term load variation.
Slack capacity is a natural target for power reduction because, in theory, it can be harvested with minimal impact on the service.
In practice, the impact of reducing slack capacity strongly depends on the service architecture dimensions listed above, which are not always visible to the cloud platform.
Fortunately, cloud operators can observe several explicit VM features, such as the presence of network-attached storage—which is associated with statefulness—that correlate with these dimensions and can serve as proxies when ground-truth information is unavailable.

\subsection{Impact Model}

Reclaiming power from a service reduces the capacity available to its workloads and
risks degrading their performance. Cloud applications are dominated by request-serving
workloads, for which this degradation manifests as dropped or timed-out requests.
To quantify the impact on a service when some of its capacity
(i.e., VMs) is turned off or throttled, we re-purpose the Conditional Value at Risk
(CVaR) measure from finance~\cite{cvar}, also known as \emph{Expected Shortfall}
(ES).
In the cloud resource setting, ES can quantify the amount of a service's load
that is at risk of being dropped due to a capacity shortfall, weighted by the observation probability of
that amount of load; the impact of a control action can be interpreted as the expected un-served
load. Formally, we define the impact of reducing service $s$'s capacity to $cap'$ as:
\begin{equation}
\label{eq:impactFunc}
    I_s(cap') = \frac{1}{load_{avg}}
    \int_{cap'}^{cap^*} (load - cap')\, PDF(load)\, \mathrm{d}\,load,
\end{equation}
where $cap^* = \sum_{w \in W_s} W_w^C$ is the service's total CPU capacity at the
onset of the power event, $cap'$ is its post-action capacity, $PDF(load)$ is the
probability density of the service's load, and $load_{avg}$ is its average load.
Intuitively, the integral accumulates the capacity shortfall $(load - cap')$ across
all load levels that exceed the reduced capacity $cap'$, weighted by how likely each
level is; normalizing by $load_{avg}$ expresses the
\textbf{expected shortfall as a fraction of average load}.

The ES model makes a simplifying assumptions on the impact of shutdowns: load on the service is assumed to be distributed among active VMs. Short-term impacts from the rerouting of active requests are not captured in this model, although they can be approximated with a per-VM penalty term weighted by utilization, as a proxy for rerouting volume.

Notably, cloud platforms have several proxies for estimating service load, such as CPU utilization and network traffic, which can be used to infer the expected distribution of future load.
An important property of the ES metric is that it approaches 1 when $cap'\to 0$ and is 0 when $cap'\ge cap^*\ge load_{max}$, making impact comparable across services.

As an example of service impact functions, we collect recent CPU usage traces in each hour of the day to discriminate between service load patterns over a day for several services. We then fit a set of common probabilistic distributions using the Fitter library ~\cite{fitter} and Scipy's gaussian\_kde method \cite{2020SciPy-NMeth} to the usage data points of each hour. The fitted distributions enable convenient PDF calculation that are differentiable -- a property that could be exploited by optimization algorithms.

\begin{figure}[h]
    \begin{subfigure}{\columnwidth}
        \centering
        \includegraphics[width=0.48\columnwidth]{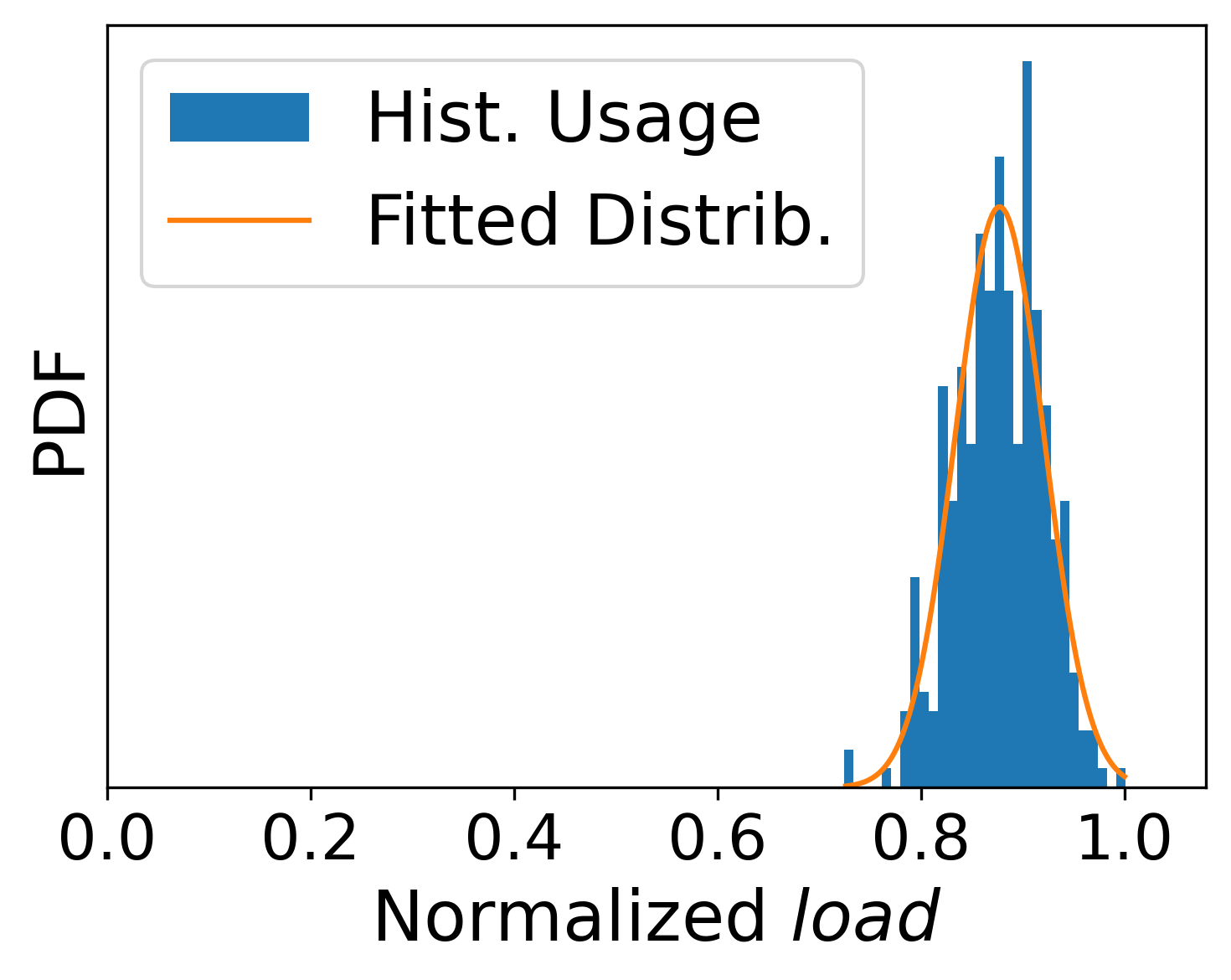}
        \includegraphics[width=0.48\columnwidth]{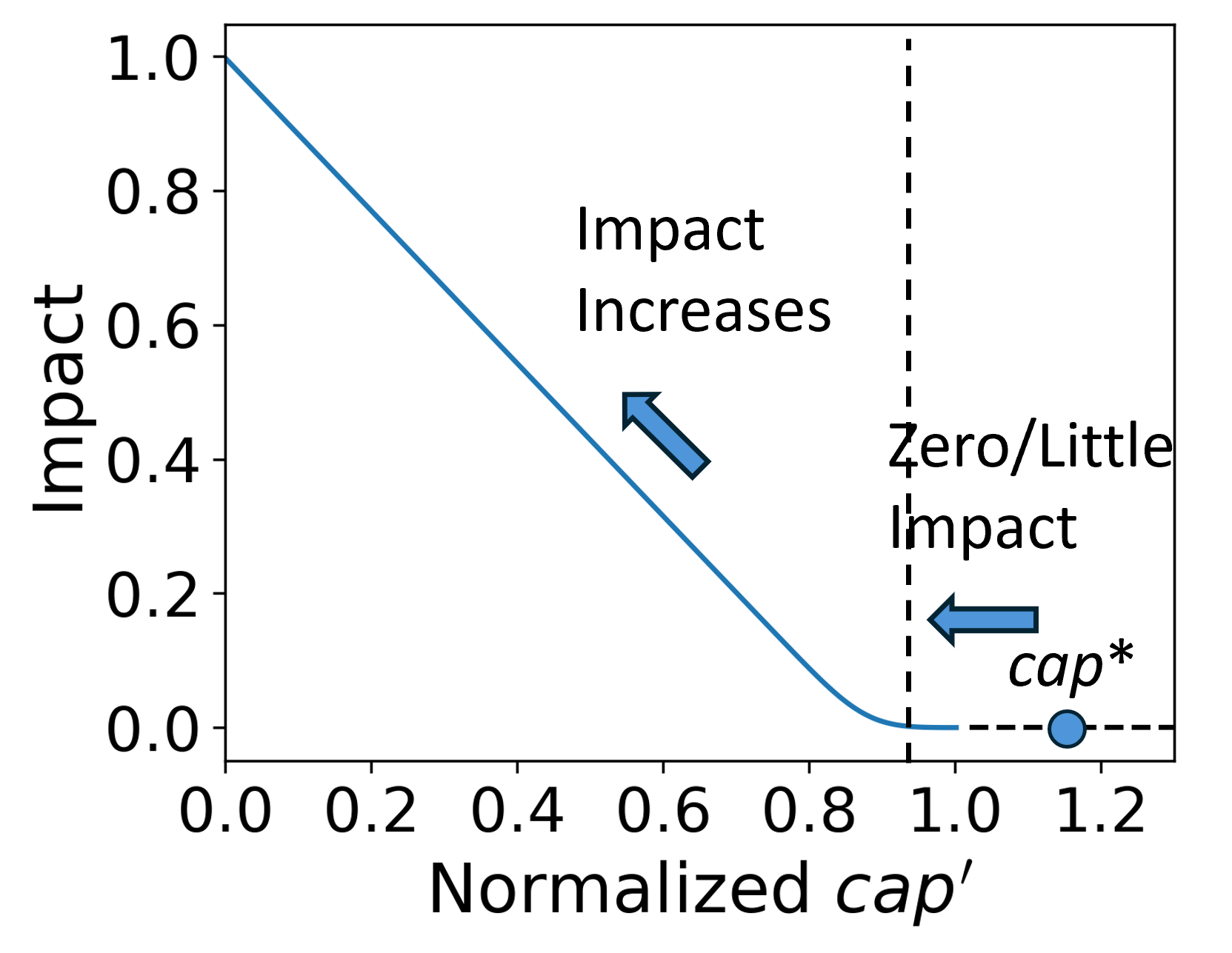}
        \caption{Service A (Best-fit distribution: normal).}
    \end{subfigure}
    \begin{subfigure}{\columnwidth}
        \centering
        \includegraphics[width=0.48\columnwidth]{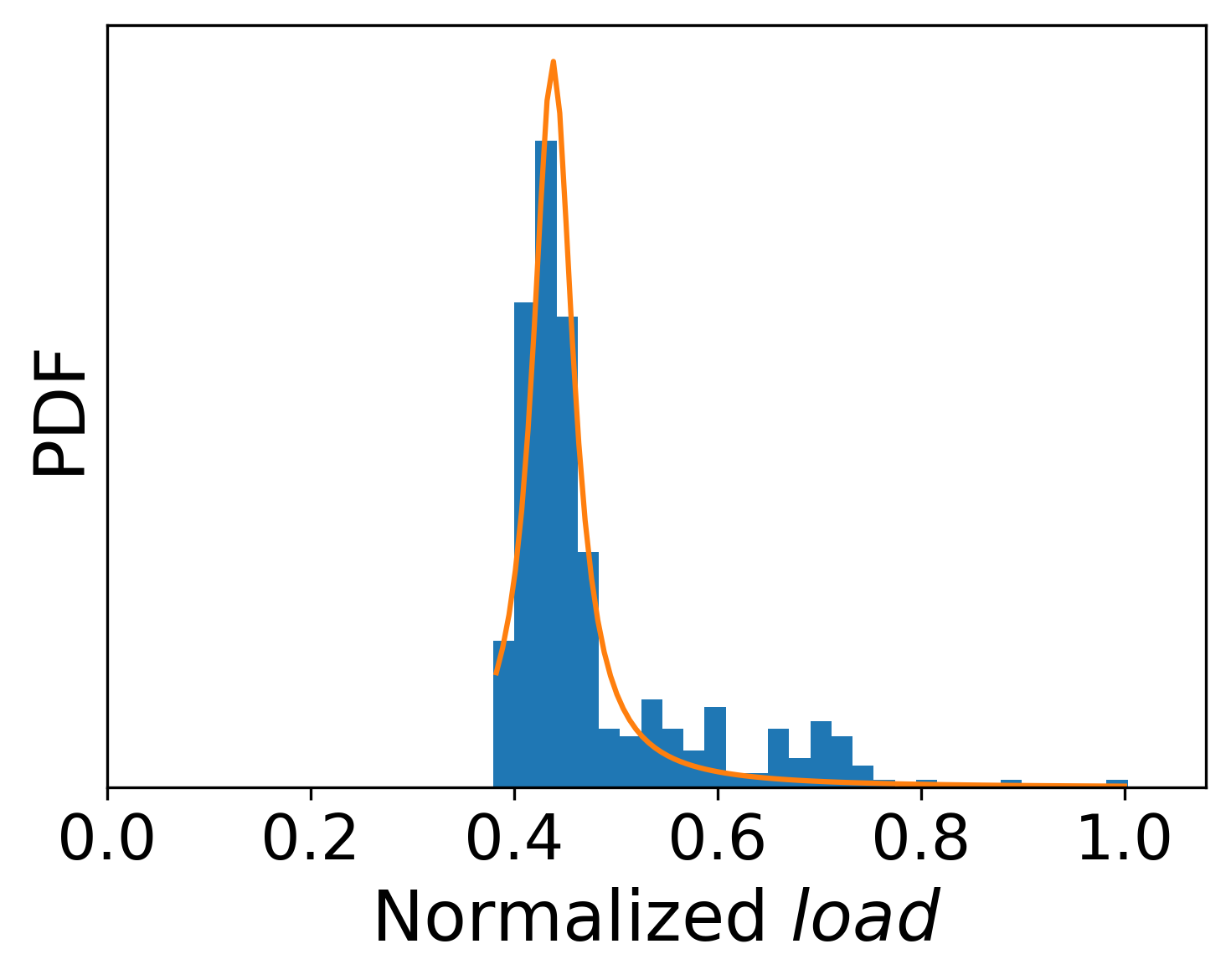}
        \includegraphics[width=0.48\columnwidth]{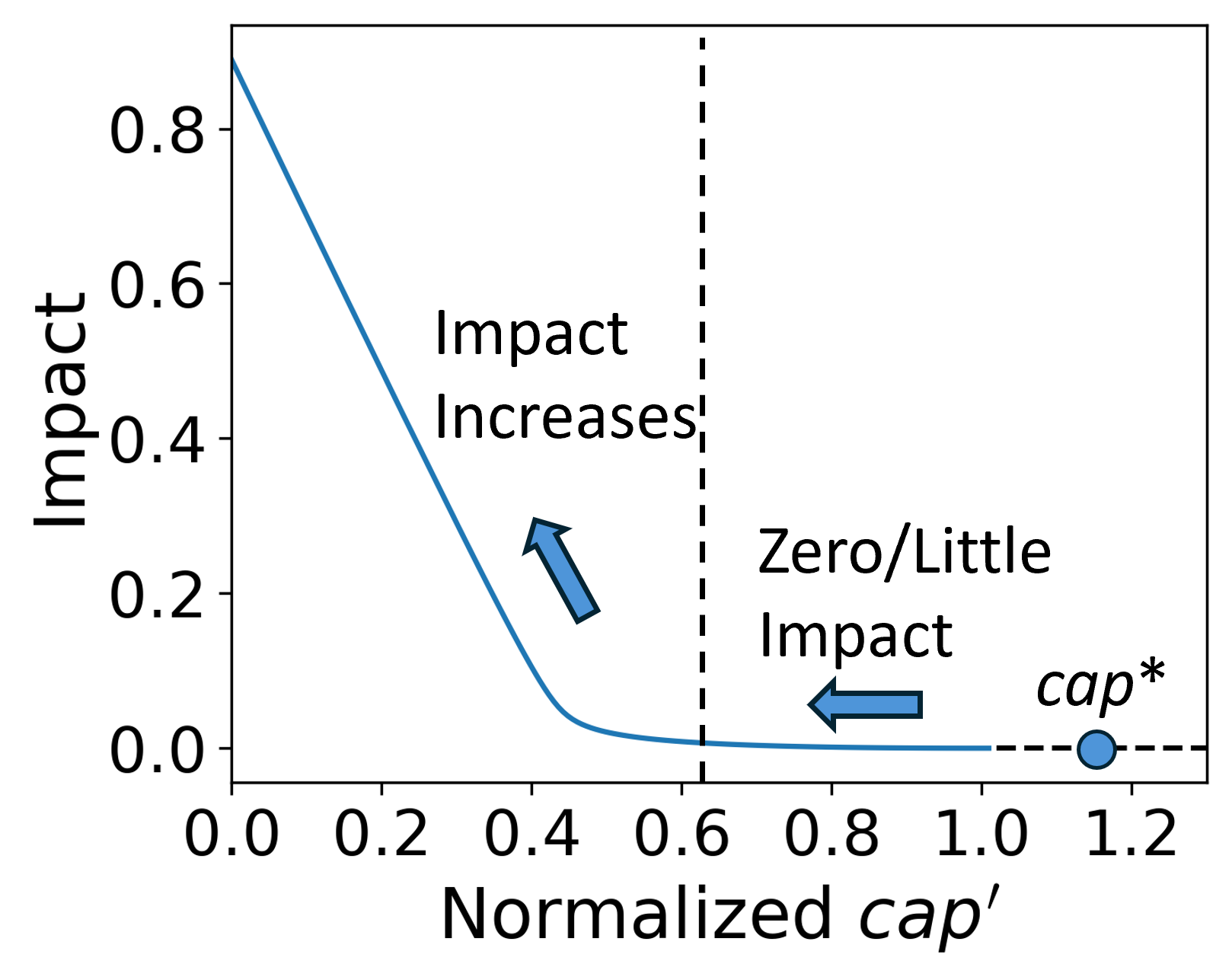}
        \caption{Service B (Best-fit distribution: Cauchy).}
    \end{subfigure}
    \centering
    \caption{Examples of historical load (normalized to Max.), fitted distributions, and ES impact functions, showing how the sensitivity to capacity reduction can vary across services.}
    \label{fig:serviceImpactExample}
    \vspace{-0.2in}
\end{figure}

Figure \ref{fig:serviceImpactExample} shows examples of service impact functions. On the left side, the historical usage of VMs corresponding to two services is represented as histograms with fitted distributions in orange. The corresponding impact calculated with (\ref{eq:impactFunc}) across a varying remaining capacity $cap'$ is shown on the right. The slack between the pre-event capacity $cap^*$ and actual load enables capacity reduction with zero or little impact, whose magnitude can be significantly different among services. For example, while keeping the expected shortfall negligible, Service B can tolerate more capacity loss than Service A. Our impact model captures this difference.

Without knowledge of application performance metrics, it is difficult to distinguish an application with over-provisioned capacity from an application where increased server utilization would degrade performance.
The impact calculation can be adjusted to assume that the service is right-sized and rescale the historical usage so that $load=1$ corresponds to the observed maximum, producing a more conservative estimate of impact.

\subsection{Total Cost of an Action}
Actions in response to an event (e.g. shut down servers) can involve multiple services. To model the overall cost (negative impact), a weight $w$ reflects service priority based on a series of properties that can be inferred from data available to the cloud platform. Table \ref{tab:priorityScore} lists a set of usable properties and corresponding relative priority ordering. The information availability of these properties varies: ownership and VM type are known, and other properties such as user-facing can be inferred through utilization patterns. The weight for service $s$ ($w_s$) is the product of the scores assigned to each property. For new services with little historical data, high priority scores can be used to avoid control interference until its properties and usage patterns become clear.

\begin{table}[h]
\caption{Service Priority Score Sheet}
\label{tab:priorityScore}
\begin{center}
\begin{tabular}{p{95pt}p{30pt}p{75pt}}
  \hline
  \textbf{Property} & \textbf{Yes} & \textbf{No}\\
  \hline
  External & 1000 & \shortstack[l]{100 (prod);\\10 (non-prod)}\\
  Spot VM & 0 & 1\\
  User-facing & 1.5 & 1\\
  Oversubscribable & 0.5 & 1\\
  External Storage & 0.5 & 1\\
  Software-redundant & 0.25 & 1\\
  \hline
\end{tabular}
\end{center}
\end{table}

The optimization objective for \sysname controllers is the total impact, or cost, of an event response:
\begin{equation}
\label{eq:totalCost}
cost=\sum_{s\in S} w_s\cdot I_s(cap_s')
\end{equation}
where $S$ denotes the set of hosted services.
The general optimization problem for determining the control actions of \sysname to minimize this cost is described in Appendix~\ref{app:opt}.

\section{Control Policy for Server Shutdown} \label{sec:controlpolicies}
\sysname control policies aim to minimize the service impact of power modulation by carefully selecting where to reduce power. Some levers, such as turning off unallocated servers or diverting new deployments, can be deployed to an extent with almost no impact to hosted services but may not reduce enough power in a timely manner for extreme scenarios such as grid emergencies.

Server base power accounts for a large fraction of overall power (Figure~\ref{fig:dc-power-breakdown}) and servers can be shut down quickly, but the impact of shutting down allocated servers is significant. Hence, the selection of servers for shutdown must be done carefully to balance power with risk. In this section we look at algorithms for shutting down allocated servers. Our study does not include power throttling, which has a limited effect at the scale of a datacenter, or VM consolidation levers, which are more suitable for cases where sufficient notice time is provided. We defer a thorough study of consolidation and other levers for future work.

\subsection{Algorithm description}

The algorithm selects the set of servers that, when shut down, will meet a specified power reduction goal.
The ES impact functions computed using Eqs.~\ref{eq:impactFunc} and \ref{eq:totalCost} guide server selection.
The challenge is that server-level impacts are not additive when services span multiple servers.
For example, the loss of any one server may represent a services' slack capacity, but losing two servers may cause an under-capacity event.
Since the marginal impact on a service increases with every reduction in service capacity, impact of each set must therefore be evaluated over the entire set.
With thousands of services distributed across thousands of servers, the number of possible server sets is combinatorial.
Despite these challenges, our evaluation shows that a simple greedy algorithm (Alg.~\ref{alg:nonEmptyServerSelPolicy}) quickly identifies sets of servers to hut down with low impact.
The algorithm iteratively selects servers from a list ordered by increasing aggregate impact per Watt.
After each selection, the shutdown cost of remaining servers hosting a VM from an affected service is recalculated and re-inserted into the sorted list.
This continuous re-evaluation (lines 13--19) is essential to avoid repeatedly drawing on the same slack capacity.

\begin{algorithm}[h]
    \renewcommand{\algorithmicrequire}{\textbf{Input:}}
    \renewcommand{\algorithmicensure}{\textbf{Output:}}
    \caption{Server Selection Policy}
    \label{alg:nonEmptyServerSelPolicy}
    \begin{algorithmic}[1]
        \REQUIRE Power reduction goal $powRedGoal$, server and VM snapshots.
        \ENSURE Actions on servers $actions$.
        \STATE $actions = \{\}$
        \STATE $expPowRed \gets 0$
        \FOR{each service $s\in S$}
        \STATE $\Delta cap_{s}\gets 0$
        \STATE $\Delta cap_{s,lastEval}\gets 0$
        \ENDFOR
        \FOR{each server $i\in servers$}
        \STATE $score_i\gets cost_i/pow_i$
        \STATE Push $score_i$ with snapshot info and $\Delta cap_{s,lastEval}$ into a minimum heap based on the score.
        \ENDFOR
        \WHILE{$expPowRed<powRedGoal$}
            \STATE Pop server $i$ info from the heap.
            \IF{$\Delta cap_s=\Delta cap_{s,lastEval}$ for each service $s$ running on $i$}
            \STATE $actions\gets actions\cup \{i\}$
            \STATE $expPowRed\gets expPowRed+pow_i$
            \STATE $\Delta cap_s \gets \Delta cap_s+cap_{s,i}$
            \ELSE
            \STATE Recalculate $score_i$ based on current $\{\Delta cap_{s}\}$.
            \STATE Push the updated $score_i$ and info back to the heap.
            \ENDIF
        \ENDWHILE
    \end{algorithmic}
\end{algorithm}

This algorithm is designed for \sysname local controllers, which can calculate service impact and power using datacenter telemetry to produce a sorted list of servers and cumulative power reduction. The runtime evaluation in Section~\ref{sec:peakshavingresults}) shows that computation takes seconds, enabling the list to stay consistent with an ever-changing datacenter and service state.

\subsection{Evaluation}
\label{sec:caseStudy}

\paragraph{Experiment Setup}
We compare three policies in \simulatorname using production datacenter trace consisting of:
\begin{itemize}
\item Service load history: 30 days of historical aggregate CPU utilization used as a proxy for service load.
\item Power telemetry: 1 hour of power telemetry attributed to VMs and other workloads.
\item VM traces: 1 hour of VM allocation records.
\end{itemize}

Given a datacenter-wide power reduction goal, these policies first shut down empty servers until the limit of minimum number of empty servers for each cluster is reached, and they differ in the selection of allocated servers to shut down:
\begin{itemize}
    \item \textbf{AscUtil}: shut down servers in ascending order of server utilization.
    \item \textbf{AscScore-OnePass}: shut down servers in ascending order of score (cost estimation). The scores are calculated only once at the start.
    \item \textbf{AscScore-ReEval}: shut down servers in ascending order of score, and the score is re-evaluated when servers with overlap in running services have been selected (Alg. \ref{alg:nonEmptyServerSelPolicy}).
\end{itemize}
The impact of each service is weighed equally regardless of the number of underlying VMs, and scaled with priority scores computed according to VM properties as detailed in~\ref{app:priorityscoresheet}.

\paragraph{Results} \label{sec:peakshavingresults}

\begin{figure}[h]
    \begin{subfigure}{0.48\columnwidth}
        \includegraphics[width=\columnwidth]{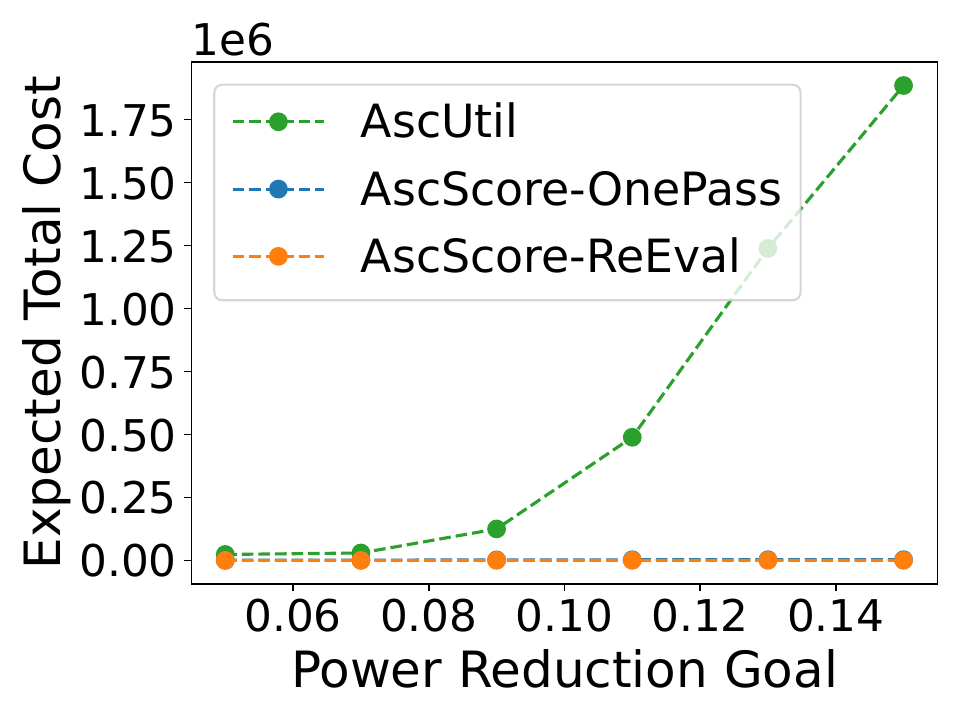}
        \caption{All Three Policies.}
    \end{subfigure}
    \begin{subfigure}{0.48\columnwidth}
        \includegraphics[width=\columnwidth]{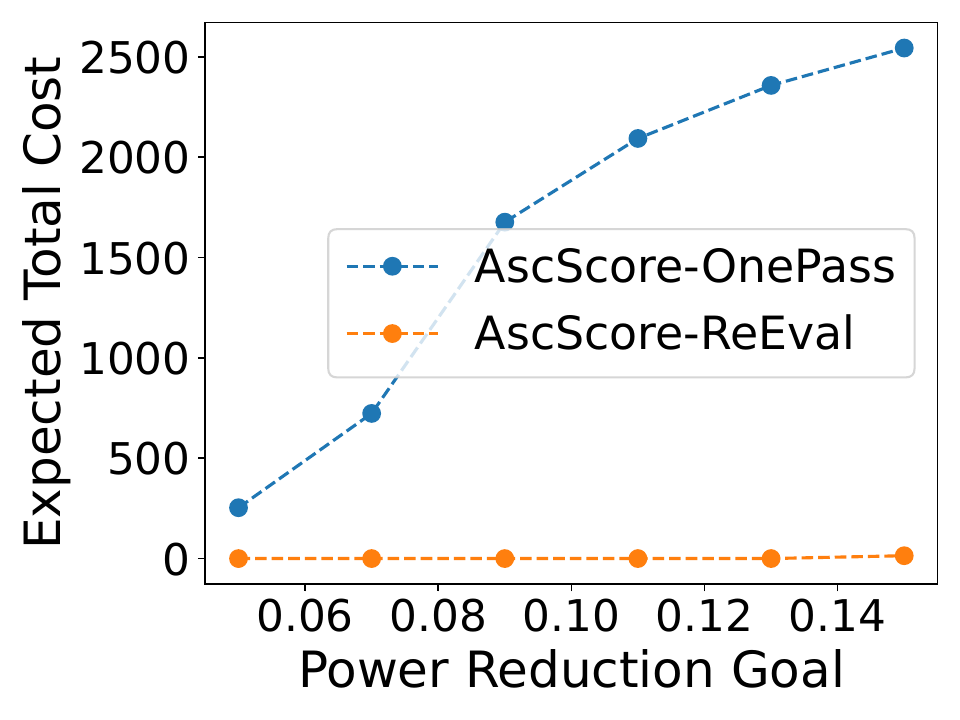}
        \caption{AscScore Variants.}
    \end{subfigure}
    \centering
    \caption{Agg. impact of various power reduction goals.}
    \label{fig:costVsPowerReduction_peakShaving}
    \vspace{-0.15in}
\end{figure}

\begin{figure}[h]
    \begin{subfigure}{0.48\columnwidth}
        \includegraphics[width=\columnwidth]{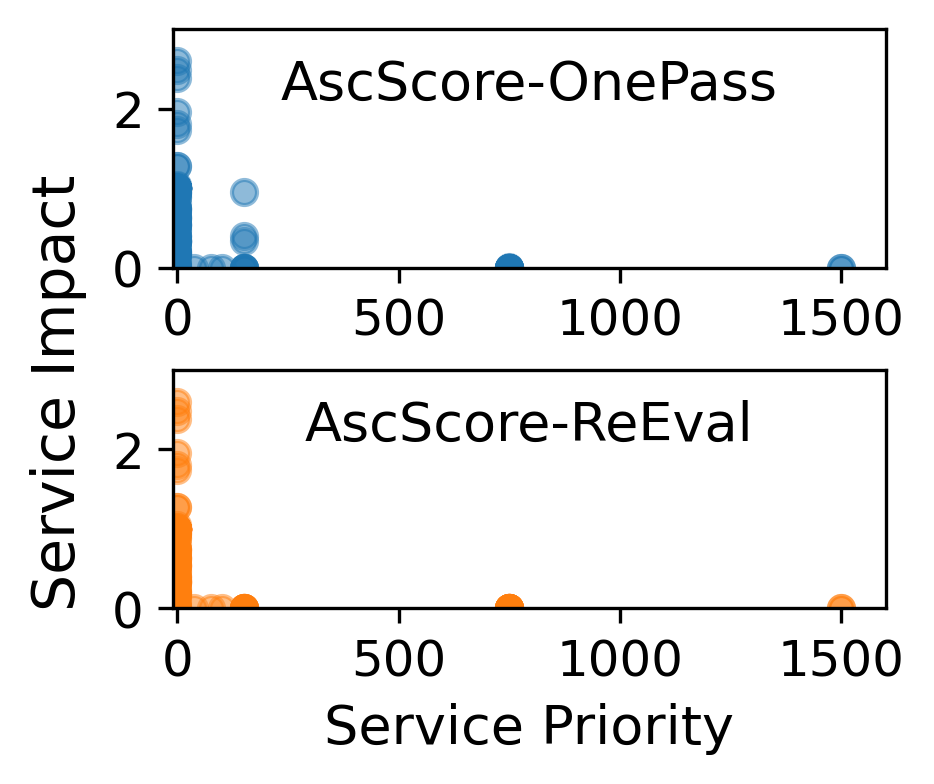}
        \caption{$powRedGoal=0.05$.}
    \end{subfigure}
    \begin{subfigure}{0.48\columnwidth}
        \includegraphics[width=\columnwidth]{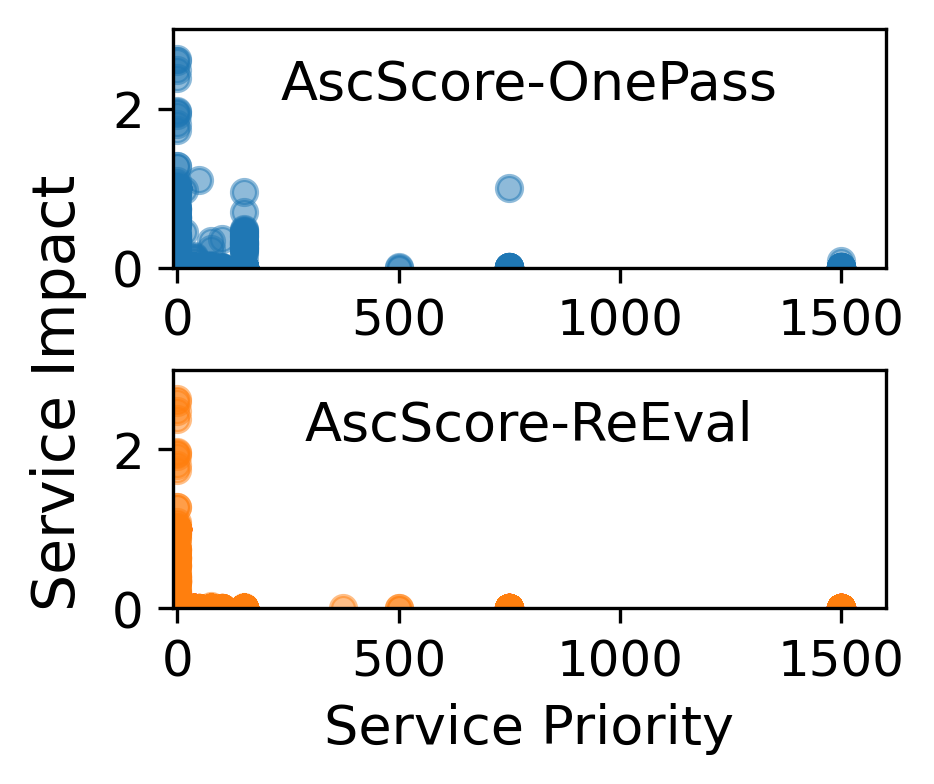}
        \caption{$powRedGoal=0.15$.}
    \end{subfigure}
    \centering
    \caption{Per-service impacts across algorithms and reduction goals. AscScore-ReEval impacts fewer high-priority services.}
    \label{fig:costBreakdown_peakShaving}

\end{figure}

Figure \ref{fig:costVsPowerReduction_peakShaving} compares different policies' expected total cost, aggregated across all services at various power reduction targets.
The first 5\% reduction is achieved by shutting down excess unallocated servers without violating buffer constraints.
AscUtil, which only considers server metrics, quickly incurs a large cost as required power reduction increases. As a reference, for a user-facing external service, an expected capacity shortfall equal to average load incurs a cost of 1500, so the total cost implies an equivalent of over 1000 such services suffer from capacity shortfall for a 15\% power reduction.

As shown in Figure~\ref{fig:costBreakdown_peakShaving}, AscScore variants select servers that mainly host spot VMs (0 priority) or run services that have sufficient slack capacity, significantly reducing the impact. When comparing AscScore-OnePass and AscScore-ReEval, only using the initial server scores results in many services seeing capacity reduction beyond their slack (reflected in non-zero impacts), while re-evaluating the score keeps the cost near zero even at 15\% power reduction.

\begin{figure}[h]
    \centering
    \includegraphics[width=0.8\columnwidth]{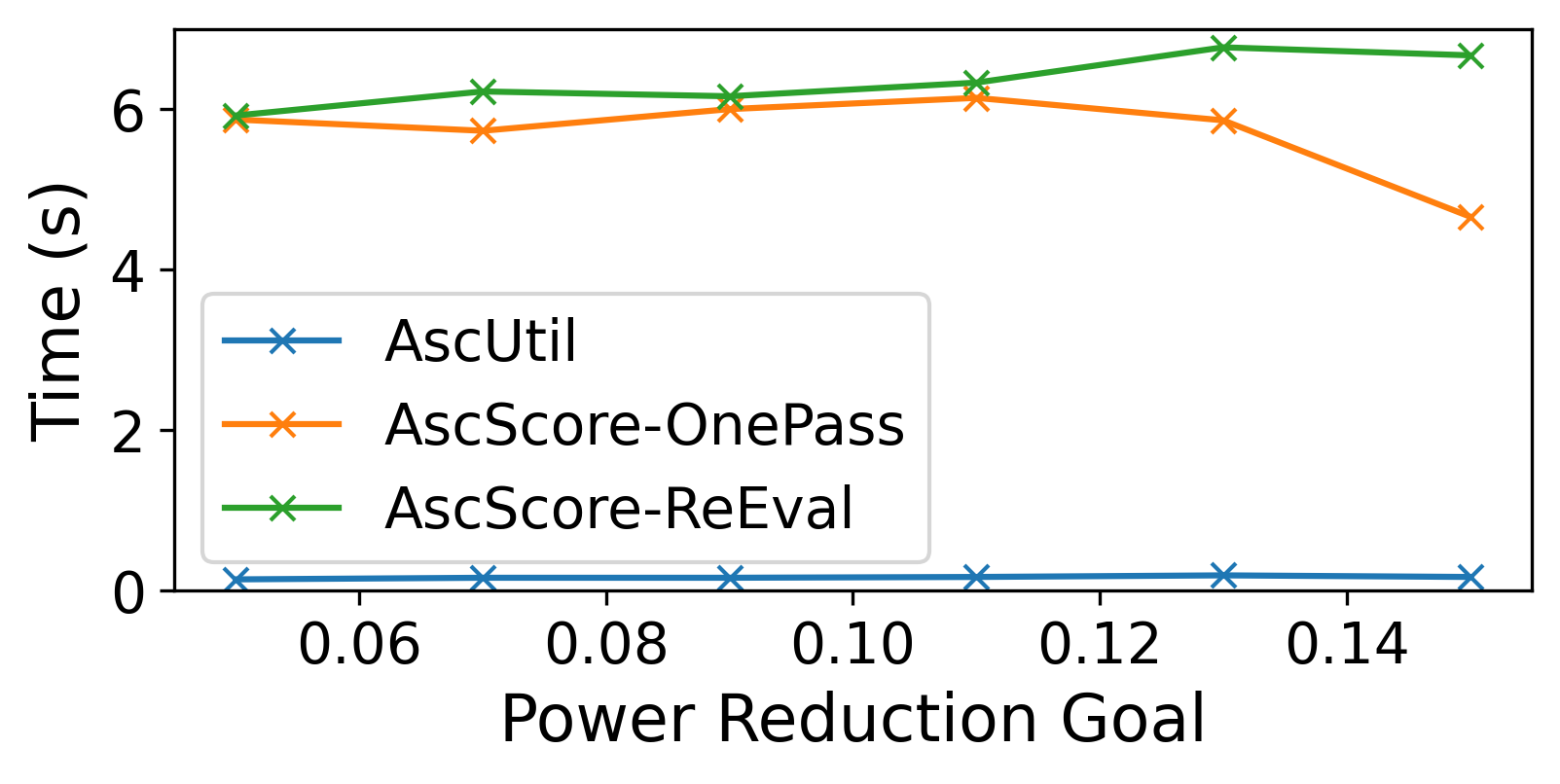}
    \caption{Computation time of control policies.}
    \label{fig:algoRunningTime_peakShaving}

\end{figure}

We also measure the amount of time required to compute actions for each policy to see if they can keep up with a dynamic datacenter state.
Figure \ref{fig:algoRunningTime_peakShaving} shows the single-thread running time of different control policies for the initial response, where most of the control actions are decided. For this sampled datacenter with about 5,000 non-empty (allocated) servers and 10,000 services, evaluating server scores takes 4.5--6 seconds (difference between AscUtil and AscScore-OnePass). The re-evaluation with AscScore-ReEval adds another 0.1--2 seconds, which increases with number of non-empty servers to select. These running times do not significantly increase staleness of control decisions relative to existing telemetry and control-propagation delays.

\section{Summary}
With growing power demands, modulation is an increasingly important requirement for datacenters.

In this paper, we introduce \sysname, a general power management system that coordinates hardware and software levers to meet power targets with minimal impact on workloads.

We develop \simulatorname, a simulator for testing power management policies using high-resolution datacenter power traces.

We also introduce a risk-based service impact model to quantify the impact of throttling or shutting down VMs.

Our experiments show that \sysname control policies can leverage any available slack capacity across services, and avoid impact to high-priority services while considering thousands of servers and services.
Overall, these results suggest that coordinated, impact-aware policies can enable datacenters to achieve modest power reduction goals while preserving service quality.

\sysname is built to support a wide set of scenarios; tailoring policies for each class of scenarios is a rich area for further study.
Many open problems exist, including power reduction of cloud storage systems, and algorithms for navigating the rich action space and generating cohesive and optimized control policies.
AI workloads, in particular, offer opportunities for geo-distribution and local modulation.
We hope our work informs more extensive research into datacenter power flexibility.

\clearpage
\bibliographystyle{ACM-Reference-Format}
\bibliography{refs}

\appendix
\section{Evaluated workloads}\label{app:workloads}

The CPU frequency sensitivity study was conducted on the following ten request-serving applications.
All applications run in CPU VMs without attached accelerators.
Across different applications, request paths span compute, memory, and storage operations.
Most tested applications resemble typical cloud workloads; Go workloads were customized to stress specific resources.

\noindent\textbf{PostgreSQL.} A transactional TPC-B-like workload generated with PostgreSQL's pgbench~\cite{postgresql2026postgresql}, with a 3.09~ms p99 SLO.\par
\noindent\textbf{Go HTTP (compute).} A custom service built with Go's net/http package~\cite{go2026nethttp} whose request handler runs a tight integer-mixing loop, with an 8.5~ms p99 SLO.\par
\noindent\textbf{TensorFlow Serving.} A machine-learning inference service implemented with TensorFlow Serving~\cite{tensorflow2026tensorflow}, with a 62~ms p99 SLO.\par
\noindent\textbf{Go HTTP (pointer chasing).} A custom net/http service mode~\cite{go2026nethttp} that pointer-chases through a randomized 1~GiB shared working set, with a 22~ms p99 SLO.\par
\noindent\textbf{Keycloak.} A JVM authentication and token service implemented with Keycloak~\cite{keycloak2026keycloak}, with a 40~ms p99 SLO.\par
\noindent\textbf{Neo4j.} A JVM-based graph-database query workload implemented with Neo4j~\cite{neo4j2026neo4j}, with a 2.5~ms p99 SLO.\par
\noindent\textbf{memcached.} The memcached in-memory key-value service~\cite{memcached2026memcached} uses a 5~ms p99 threshold.\par
\noindent\textbf{imgproxy.} The imgproxy image-resizing service~\cite{imgproxy2026imgproxy}, uses a 75~ms p99 SLO.\par
\noindent\textbf{MinIO.} The MinIO S3-compatible small-object storage service~\cite{minio2026minio} uses a 2~ms p99 threshold.\par
\noindent\textbf{HashiCorp Vault.} A Vault transit-encryption workload~\cite{hashicorp2026vault} with an 8~ms p99 SLO.

\section{Optimization problem formulation} \label{app:opt}

The optimization problem in this section, when solved, determines the combination of the following actions that minimize the impact on the workload while meeting the power reduction target:
\begin{itemize}
\item Worker state $\alpha_w$ for worker $w$, a continuous value in $[0,1]$ that scales the CPU frequency of the corresponding cores from $[0,CPU_{max}]$
\item Server state $\beta_k$ for server $k$, and integer which can be a binary on/off indicator or have intermediate low-power states.
\item Worker-server binary allocation matrix $A$, where $A_{w,k} = 1$ if worker $w$ is allocated to server $k$
\end{itemize}

\subsection{Worker and server models}
Workers (i.e., VMs) $w \in W$ are modeled with an implicit average utilization that determines their dynamic power.
This power scales with the frequency of the cores they run on.

Servers $k \in K$

have an associated baseline power $b_k$ that represents the idle power of the CPU, memory, network card, and other components.
The assignment of workers to servers is expressed as a binary allocation matrix $A$, where $A_{w,k} = 1$ if worker $w$ is hosted by server $k$.

\subsection{Power model}

Datacenter power has two components: a baseline power representing static power consumed by components such as network switches and control plane hardware, and a dynamic power that is affected by CPU throttling.
The dynamic power can be calculated bottom-up using the power consumption of each worker and the base (idle) power of each server.
Each worker's dynamic power is modeled with a function $f$ of the CPU frequency and server state:

\begin{align}
    f: \alpha_w, \beta_k \rightarrow \text{worker $w$ dynamic power}
\end{align}

where $f(\alpha_w, \beta_k) = 0$ if $A_{w,k} = 0$ (i.e., $w$ is not allocated to $k$), or $\beta_k = 0$ (server $k$ is off).

Server $k$ power ($P_k$) is a function of server base power and the dynamic power of hosted workers, for example as in Figure~\ref{fig:serverpower_freq_throughput}, captured in the following equations and constraints:
\begin{align}
P_w = f(\alpha_w, \beta_k) \label{eq-workerpower}\\
P_k = \Big(\sum_{w} A_{w,k} P_w \Big) + b_k \beta_k \label{eq-serverpower} \\
0 \leq A_{w,k} \beta_w \leq \beta_k  \label{eq-alphabeta}
\end{align}

The last constraint ensures workers hosted by $k$ are off if the server is off. This formulation assumes base power is the same at any CPU frequency, which might depend on processor states.
It also ignores the impact of setting the entire core's frequency vs. setting each worker's frequency dynamically; also assumes that individual workers can be throttled on a multi-tenant server.

The datacenter power depends on the baseline power $B$, the server power, and power usage effectiveness ($PUE$), a factor that captures the additional power consumed by power infrastructure and cooling systems:
\begin{equation}
    PUE \Big(B + \sum_{\text{server $k$}} P_{k}\Big)
\end{equation}

\subsection{Putting it together}

Given the datacenter target power $P_{target}$, datacenter base power ($B$) and $PUE$, the set of functions representing worker dynamic power ($f$), allocation matrix $A$, and hosted service impact functions $(I_s)$ and weights $w_s$, determine the set of actions ($\alpha_w$, $\beta_k$) that brings datacenter power to $P_{target}$ while minimizing impact on the hosted services.

\begin{align}
\min_{\alpha_w, \beta_k} \sum_{s \in S} w_s \cdot I_s \\
\text{Constrained on:} \nonumber\\
PUE \Big(B + \sum_{k \in K} P_{k}\Big) \leq P_{target} \nonumber\\
\text{Eqs.~\ref{eq-workerpower}, \ref{eq-serverpower}, \ref{eq-alphabeta}} \nonumber\\
0 <= \alpha_w <= 1, ~~\forall w \in W \\
\beta_k ~\in ~\{0,1\}, ~~\forall k \in K
\end{align}

\section{Service priority configuration} \label{app:priorityscoresheet}

\begin{table}[!htbp]
\centering
\caption{Service Priority Score Sheet}
\label{tab:priorityScorevals}
\begin{tabular}{p{95pt}p{30pt}p{75pt}}
  \hline
  \textbf{Property} & \textbf{Yes} & \textbf{No}\\
  \hline
  External & 1000 & \shortstack[l]{100 (prod);\\10 (non-prod)}\\
  Spot VM & 0 & 1\\
  User-facing & 1.5 & 1\\
  Oversubscribable & 0.5 & 1\\
  External Storage & 0.5 & 1\\
  Software-redundant & 0.25 & 1\\
  \hline
\end{tabular}
\end{table}

Table~\ref{tab:priorityScorevals} describes the set of weight factors used to assign service priority for datacenter shedding experiments in Section~\ref{sec:controlpolicies}.

\clearpage

\end{document}